\documentclass[fleqn,usenatbib]{mnras}

\usepackage{newtxtext,newtxmath}
\usepackage[T1]{fontenc}
\usepackage{listings}
\DeclareRobustCommand{\VAN}[3]{#2}
\let\VANthebibliography\thebibliography
\def\thebibliography{\DeclareRobustCommand{\VAN}[3]{##3}\VANthebibliography}

\usepackage{graphicx}	% Including figure files
\usepackage{amsmath}	% Advanced maths commands

\title[First on-sky fringes for CHARIOT]{Stellar interferometry with a K-band integrated optics beam combiner instrument at the CHARA Array - I. First on-sky fringes for CHARIOT}

\author[A.N. Dinkelaker et al.]{Aline N. Dinkelaker,$^{1}$\thanks{Present address: Corning Optical Communications GmbH \& Co. KG, Walther-Nernst-Str. 5, 12489 Berlin, Germany}
Kévin Barjot,$^{2}$\thanks{E-mail: barjot@ph1.uni-koeln.de}
Nic Scott,$^{3}$
Aurélien Benoît,$^{4}$
Narsireddy Anugu,$^{3}$
Alyssa V. Mayer,$^{5}$
\newauthor
Jacopo Siliprandi,$^{4}$ % jacopo.siliprandi@gmail.com
Robert R. Thomson,$^{4}$
Kalaga Madhav,$^{1}$
Gail Schaefer,$^{3}$
Lucas Labadie$^{2}$\thanks{E-mail: labadie@ph1.uni-koeln.de}
\\
\\
$^{1}$Leibniz-Institut für Astrophysik Potsdam, An der Sternwarte 16, 14482 Potsdam, Germany\\
$^{2}$I. Physikalisches Institut, Universität zu Köln, Zülpicher Straße 77, 50937 Köln, Germany\\
$^{3}$The CHARA Array of Georgia State University, Mount Wilson Observatory, Mount Wilson, CA 91023, USA\\
$^{4}$Institute of Photonics and Quantum Sciences, Heriot-Watt University, Edinburgh EH14 4AS, UK \\
$^{5}$Humboldt-Universität zu Berlin, Department of Physics, Newtonstr. 15, 12489 Berlin, Germany\\
}

\date{Accepted XXX. Received YYY; in original form ZZZ}

\pubyear{\the\year{}}

\begin{document}
\label{firstpage}
\pagerange{\pageref{firstpage}--\pageref{lastpage}}
\maketitle

\begin{abstract}
We present the instrument description and the first on-sky observations for the new CHara ARray Integrated Optics Testbench (CHARIOT) at the CHARA Array. 
CHARIOT is a 2-telescope proof-of-concept instrument that utilises a fibre-connectorised beam combiner fabricated by ultrafast laser-inscription in Infrasil quartz glass, which presents low-OH contamination and high transparency over the whole K-band. 
With its fibre-connected architecture, CHARIOT is designed as a plug-and-play testbed instrument to test and validate fibred astrophotonic components on-sky.
At this point, CHARIOT is able to measure raw fringe squared visibilities with a sub-1\% precision for stars with K-mag $\lesssim$ 2. Under poor to moderate seeing conditions, the current limiting magnitude for fringe detection is K-mag = 5.212 (HD~210824). These results demonstrate that CHARIOT's performance is already comparable to the former JouFLU instrument. 
With additional improvements, the limiting magnitude could be extended and the precision for fainter stars improved. The successful on-sky demonstration paves the way for a new high precision K-band beam combiner instrument at the CHARA array.
\end{abstract}


%%%%%%%%%%%%%%%%%%%%%%%%%%%%%%%%%%%%%%%%%%%%%%%%%%

%%%%%%%%%%%%%%%%% BODY OF PAPER %%%%%%%%%%%%%%%%%%

%%%%%%%%%%%%%%%%%%%%%%%%%%%%%%%%%%%%%%%%%%%%%%%%%%%%%%%%%%%%%%%%%%%%%%
\section{Introduction} 

Long-baseline interferometry at near-infrared wavelengths enables observations of astrophysical objects with an angular resolution $\lambda/B$ at the milliarcsecond level. Currently, this is achievable with the hectometric baselines of the VLTI in Chile \citep{Woillez2018} and the CHARA Array in the US \citep{Brummelaar:2005}. In the context of this work, two science cases can benefit from the angular resolution and sensitivity reachable, for instance, at the CHARA Array. 

Firstly, the faint emission from hot ($\sim$1000--1500\,K, \citet{Kral:2017}) exozodiacal dust around early-type main-sequence stars originates from dust concentrated in the inner few astronomical units of the stellar environment. The spatial extent of these hot debris disks can be accurately measured through spatially resolved measurements. Furthermore, high-angular-resolution observations can help us constrain the level of hot-dust near-infrared excess in these systems down to a few percent, which can be otherwise difficult to detect from a classical spectral energy distribution analysis \citep{Nunez:2017b,Kirchschlager2018}. The detailed multi-wavelength characterisation of these exozodiacal disks is highly relevant not only for a better understanding of the processes of planet formation, but also for understanding how this disk-extended emission could hamper the search of small-size inner exoplanets. 
Secondly, with the upcoming delivery of the DR4-release, the GAIA mission \citep{Gaia2016} will have revealed a large number of binary systems with separations $<$0.1$^{\prime\prime}$ through the astrometry of the system photocentre. For these spatially unresolved GAIA binaries, long-baseline interferometry is ideally suited to directly measure the binary separation, which breaks the flux ratio/separation degeneracy and allows us to measure the dynamical mass of the systems \citep{Kraus:2022}.

The heart of an interferometric instrument is the beam combination unit that coherently interferes the beams from several telescopes. In the context of astronomical instrumentation, photonic-based components \citep{Norris:2019,Minardi:2021,Labadie:2022, Jovanovic:2023, Dinkelaker:2024} have shown very successful applications in the field of long-baseline interferometry. 
These small-scale devices reduce the instrument footprint, while their integrated, monolithic nature reduces the impact of environmental changes such as vibration and thermal variations, the latter of which is of particular importance for non-cryogenic instruments. 
Furthermore, the single-mode nature of the waveguides in either fibre couplers or integrated beam combiners can boost the interferometric precision to better than 1\% \citep{Foresto:1997,Anugu:2020}. 
FLUOR \citep{Foresto:2003} was the first single-mode fibre-based beam combiner instrument for interferometry.
From 2002, it was located at the CHARA Array on Mount Wilson to combine two of the six existing telescopes. The JouFLU upgrade \citep{Lhome:2012, Scott:2013, Scott:2014} used the fibre beam combiner MONA based on fluoride glasses to improve the throughput in the K-band. With JouFLU, sub-percent precision of the squared visibility was obtained for stars with magnitudes K$\,\lesssim 3$ \citep{Nunez:2017b}, with a limiting magnitude K\,$\sim$4.5 for fringe detection. 
FLUOR and JouFLU delivered high-precision measurements on stellar diameters and identified sources with faint circumstellar exozodiacal dust \citep{Absil:2006,diFolco:2007,Akeson:2009,Absil:2013,Nunez:2017b,Ertel:2025}.

Following the decommissioning of JouFLU, the NAIR/APREXIS project in collaboration with the CHARA Array aims at the development of CHARIOT (CHara ARray Integrated Optics Testbench), which exploits the FLUOR/JouFLU legacy to build a new high-precision K-band beam combiner instrument, complementing the instrument suite at the CHARA Array. An additional goal of CHARIOT is to offer a platform to test photonic interferometric components with plug-and-play capabilities. This paper describes the undertaken effort to develop CHARIOT and its first results and is structured as follows. The CHARIOT instrument is described in Section 2, which includes the setup, data acquisition, and data analysis process. Section 3 presents the experimental results of CHARIOT, both from the laboratory and during several on-sky observation nights. The paper concludes in Section 4.

%%%%%%%%%%%%%%%%%%%%%%%%%%%%%%%%%%%%%%%%%%%%%%%%%%%%%%%%%%%%%%%%%%%%%%
\section{The CHARIOT Instrument}
\begin{table}
    \centering
    \caption{Overview of the current status of the CHARIOT instrument.}
    \label{tab:chariotinstr}
    \begin{tabular}{ll}
         \hline
         Properties & CHARIOT Instrument \\
         \hline
         Location & The CHARA Array\\
         Number of beams & 2\\
         Beam combiners & 2 ULI-fabricated photonic chips (Infrasil glass) \\
                       & Non-optimised interface gluing, BC-22 \\
                       & Optimised interface gluing, BC-25 \\
        Input coupling & Telecentric f-theta lenses + motorised mirror\\
        Interface & Commercial Nufern PM1950 fibres\\
                    & Loss: 0.1\,dB/m @2.15\,$\upmu$m \& 0.5\,dB/m @2.3$\,\upmu$m\\
                    & Input: FC/PC connectors ($\sim$1.5\,m)\\
                    & Output: Custom 2x2 fibre array ($\sim$2.5\,m)\\
         Output coupling & 4-lens system (F/20)\\
         Operating wavelength & K-band \\
         Filter & K': $\lambda_C = 2.152\,\upmu$m, $\Delta\lambda = 0.326\,\upmu$m\\
         Detector & C-RED One camera (J- to K-band)\\
         \hline
    \end{tabular}
\end{table}

%%%%%%%%%%%%%%%%%%%%%%%%%%%%%%%%%%%
\subsection{Astrophysical requirements}
With long-baseline interferometry, exozodiacal dust around a star can be resolved. Because incoherent flux reduces the visibility, exozodiacal dust results in a deficit in the measured visibility compared to the expected visibility of the star itself. The contribution to the flux from the dust is on the order of 1\% of starlight, leading to a visibility deficit of 2\%. To detect the deficit, squared visibilities have to be measured with precisions better than 1\%, see e.g. \cite{Ertel:2025, Kral:2017}. 

%%%%%%%%%%%%%%%%%%%%%%%%%%%%%%%%%%%
\subsection{Instrument description}
\label{sec:instru_description}
The CHARIOT instrument is installed at the CHARA Array \citep{Brummelaar:2005} of Georgia State University located at Mount Wilson Observatory. It replaces the JouFLU beam combiner with a new one based on novel astrophotonic technologies. It also benefits from the new adaptive optics (AO) systems installed on the telescopes \citep{Che:2013, tenBrummelaar:2018, AnuguAO:2020}. The objective is to improve the sensitivity of the previous JouFLU instrument and expand the science cases related to the characterisation of exozodiacal dust and stellar companions thanks to high-precision interferometry. Currently, CHARIOT integrates a proof-of-concept 2-telescope beam combiner chip (BC) to validate the ultrafast laser-inscription (ULI) fabrication technique as well as its plug-and-play philosophy, which is the aim of this paper. Ultimately, its ambition is to develop a fully integrated plug-and-play cryo-cooled nulling instrument combining up to 4 telescopes in H-/K-band.

\begin{figure*}
    \centering
    \includegraphics[width=\linewidth]{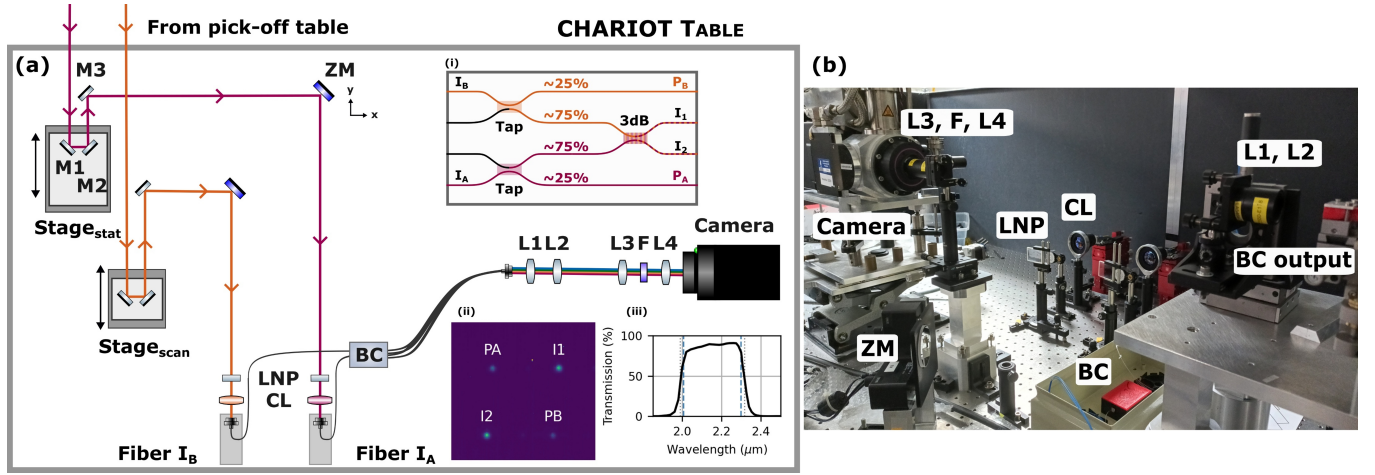}
    \caption{(a) Schematic of the CHARIOT setup in the CHARA beam combiner lab. The two beams (purple and orange) enter the CHARIOT instrument table at the top left. Using a retroreflector (M1 + M2) on 2 motorised translation stages, the path length can be coarsely adjusted by Stage$_{stat}$, and temporally scanned by the stage Stage$_{scan}$ for fringe acquisition. The beam is guided by a folding mirror (M3) to a mirror on a motorised mount (ZM), and through a birefringence compensation plate (LNP) to the fibre coupling setup. The coupling lens (CL) focuses the light beams into the two main input fibres of the BC, of which the design and the splitting ratios are shown in (i). (ii) The fibre outputs include two photometric taps (P$_A$, P$_B$) and two interferometric outputs (I$_1$, I$_2$), are arranged into a $2\times2$ fibre array, which is then imaged onto a C-RED One camera using four lenses (L1 to L4). (iii) A bandpass filter F is inserted between L3 and L4 before detection. The plot shows the filter transmission curve with its nominal bandwidth (blue dashed lines) as well as the measured (effective) system bandwidth (red solid lines). (b) The photo on the right shows part of the optical table, with some of the visible components annotated.}
    \label{fig:setup}
\end{figure*}

The CHARIOT optical setup is aligned and characterised (see Section~\ref{sec:stscharac}) by the six telescope simulator (STS), which is a thermal halogen source \citep{Anugu:2020}. The STS is aligned to the telescope beams on a regular basis. As this requires small adjustments of the STS optics, the flux in each beam is not necessarily constant for different days, and the flux ratio between beams can vary.
Once CHARIOT is aligned to the STS, switching between the STS and the telescope only requires repeating an optimisation of the fibre coupling and of the optical path difference (OPD) within the CHARIOT instrument. This is performed remotely by the operator via the control software.
The general features of CHARIOT are summarised in Table~\ref{tab:chariotinstr}, and a schematic of the instrument setup is shown in Fig.~\ref{fig:setup}. Also, more details can be found in \cite{Mayer:thesis} and \cite{Mayer:2024}, with some updates of the optical system since its publication.

The two paths on the CHARIOT table have been designed and aligned to compensate for differences introduced upstream in the CHARA laboratory, in order to match the OPD at a mm level. 
The goal is to match the OPD within the coherence length. With the K' bandpass filter (central wavelength $\lambda_C = 2.152 \,\upmu$m, bandwidth $\Delta\lambda = 0.326 \,\upmu$m), the expected coherence length is $L_{\text{coh}} = 14.2 \,\upmu$m. The measured coherence length for the entire CHARIOT optical bench is $L_{\text{coh, eff}} = 15.9\,\upmu$m, which includes additional filtering effects from fibres and optical components that effectively reduce the bandwidth ($\lambda_C = 2.152 \,\upmu$m, $\Delta\lambda_\text{eff} = 0.291 \,\upmu$m), see Fig.~\ref{fig:setup}(a)(iii). Note that the squared visibility estimator is linearly dependent on the bandwidth, thus errors in its estimation may cause a bias in the extracted squared visibility values (see Eq.~\ref{eq:V2} and Fig.~\ref{fig:dataprocessingsteps}).
The retroreflector (M1 + M2) of a motorised translation stage (Stage$_{stat}$, Newport UTS100PP, $100\,$mm travel range) enables fine-tuning of relative optical path shifts on a range of $\pm 10\,$cm. Once the lengths of the two beam paths --depicted in purple and orange-- match, the OPD of the orange path is varied around its centre (OPD$_{scan}$) to perform temporal acquisition of the fringes. This is achieved with a fast-scanning motorised stage (Stage$_{scan}$, Newport XMS50, $50\,$mm travel range, $300\,$mm/s maximum speed). All motorised translation stages are controlled with a Newport XPS motion controller system. Before entering the beam combiner chip, the light passes through a Lithium Niobate plate (LNP) in each beam path to compensate for any relative phase differences between the polarization states induced by the chip \citep{Lazareff:2012}. The plates' angles are adjusted by maximising the measured visibility of the fringes. Each beam is coupled into a polarization-maintaining input fibre (Nufern PM1950) of the BC using a mirror (ZM) mounted on a motorised tip/tilt mount (Zaber T-MM2) for automated injection and a custom telecentric f-theta lens (CL) (Wavelength Opto-Electronics) for focussing. A manual six-axis positioner (Luminos I6000) allows fine-tuning of the fibre position to optimise coupling.

The design (Fig.~\ref{fig:setup}~(a)(i)) of the ULI-fabricated photonic BC (see \cite{Benoit:2021, Siliprandi:2024} for a detailed description) is based on three asymmetric directional couplers: at each input, one coupler splits the light into a photometric arm and an interferometric arm, with an approximate ratio of 25:75 (see Appendix~\ref{sec:splitratio}). A third coupler then recombines the two interferometric paths evenly (3dB) to produce the two output signals. Eventually, the two photometric (P$_A$ and P$_B$) and the two interferometric (I$_1$ and I$_2$) output fibres are bundled in a 2x2 array, which is imaged on the C-RED One camera (Fig.~\ref{fig:setup}~(a)(ii)). Note that several BCs as described in \cite{Siliprandi:2024} were fabricated at the same time. In order to progress quickly when JouFLU was first transformed into CHARIOT, a first, temporary beam combiner (BC-22) was installed and tested, for which the fibre-connectorisation process (i.e. gluing of the V-grooves to the chip facets) had not yet been optimised. This resulted in reduced throughput. In early July 2025, the beam combiner component was updated with a second chip (BC-25), for which an optimised gluing procedure had been applied, improving the throughput by around a factor of ten. For both inputs, the total loss is $\sim$3.9\,dB (total throughput $\sim$40\%). Compared with the insertion loss of $\sim$1.1\,dB for the chip itself, the commercial fibres significantly increase the global BC losses. Swapping the two BCs was quick and easy thanks to the plug-and-play attribute of the fibre interfaces. Including the remaining adjustments (e.g. the LNP angles), the changeover was completed within a few hours, demonstrating the potential of CHARIOT as a testbench for fibred photonic component development.

In April 2024, the imaging optics were adapted from the former JouFLU NICMOS camera aperture of F/8 to the new C-RED One camera aperture of F/20, which is obtained by a cold baffle (entrance aperture) inside the camera system. In the first version of the design, a two-lens system was chosen but was affected by vignetting, thus inducing a loss of off-axis light. To still obtain usable signals during the observation nights of May 2024, the imaged flux was optimised by centring two (or in some cases one) interferometric outputs, removing the photometric outputs completely from the effective field of view. In a second step, a four-lens system was integrated in October 2024 which allowed imaging the unobstructed flux from all four outputs. This final optics system consists of the following lenses: L1 with $f = 50\,$mm (Thorlabs AL72550-E1), L2 with $f = 500\,$mm (Thorlabs LA5464-D), L3 with $f = 75\,$mm (Thorlabs ACA254-75-D) and L4 with $f = 50\,$mm (Thorlabs LA5763-D). From the C-RED One camera, the four outputs are read out in synchronisation with the translation stage scan to produce interferometric data.

Like other instruments at the CHARA Array (MIRC-X, MYSTIC, Silmaril), CHARIOT uses the C-RED One camera, but with custom specifications and modifications of the wavelength filters and apertures. For comparison, the C-RED One camera of the MIRC-X instrument is described in \cite{Anugu:2020, Lanthermann:2019, Lanthermann:2018}. Table~\ref{tab:cred} summarises the main camera specification. The built-in filters have been modified to extend the camera bandwidth from H- to K-band and a F/20 entrance aperture was installed to decrease the background noise detected by the camera. The characterisation of this specific C-RED One camera (before modification) can be found in \cite{credone:2023}.
\begin{table}
    \centering
    \caption{C-RED One camera in CHARIOT: specifications.}
    \label{tab:cred}
    \begin{tabular}{ll}
         \hline
         Parameter & Description\\
         \hline
          Filter transmission & $1 \, \upmu$m - $2.4 \, \upmu$m \\
          Sensor size & $320\times 256$ pixels with $24 \, \upmu$m each \\
          Sensor temperature & $\sim 80\,$K\\
          APD gain & $2-100$ (recommended range)\\
          Readout mode & CDS (typical) \\
          Readout noise & $< 1 e^-$/ frame / pixel rms\\
          &for APD gain $> 40$ (in CDS mode) \\
          System gain & $1.9 \,e^-$/ADU \\
         \hline
    \end{tabular}
\end{table}

For CHARIOT, correlated double sampling (CDS) is used as the default readout mode. The images provide the counts in analog-to-digital units (ADU), which is converted from electrons via the APD-Gain $G_{APD}$ and the systems gain. For more details on the camera as well as noise characterisation, see the Appendix~\ref{sec:camnoise}. Moreover, two behaviours are to be noted. First, the pulse tube cooling system was introducing vibrations in the fringe signal (a similar problem has been indicated in \cite{Lanthermann:2018}) which could be mitigated by subtracting a background pixel from the signal pixel (since October 2024). Secondly, there are also indications of a non-linear detector response with respect to light intensity, which could not yet be fully investigated. As a precaution to avoid asymmetric fringe signals, we aim to stay in the lower ADU range (up to around $10^4$ ADU) by adjusting the APD gain accordingly.

%%%%%%%%%%%%%%%%%%%%%%%%%%%%%%%%%%%
\subsection{Data acquisition}
\label{sec:daq}

\begin{figure*}
    \centering
    \includegraphics[width=\linewidth]{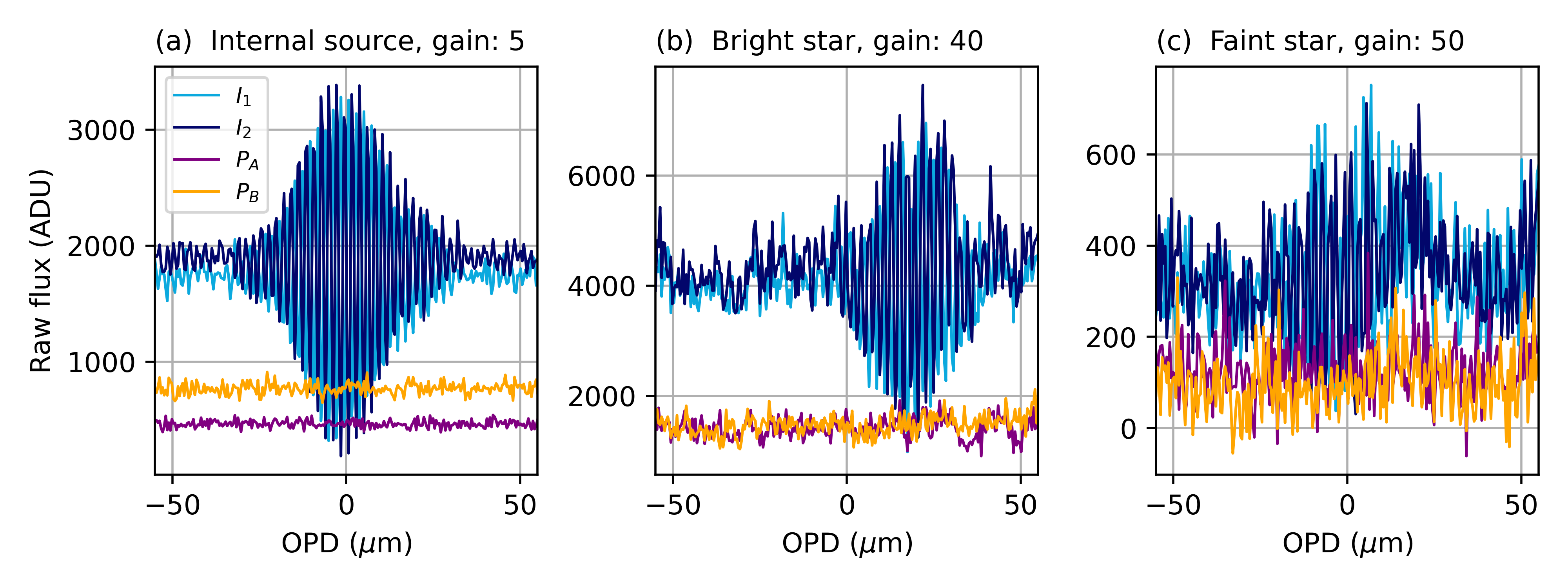}
    \includegraphics[width=\linewidth]{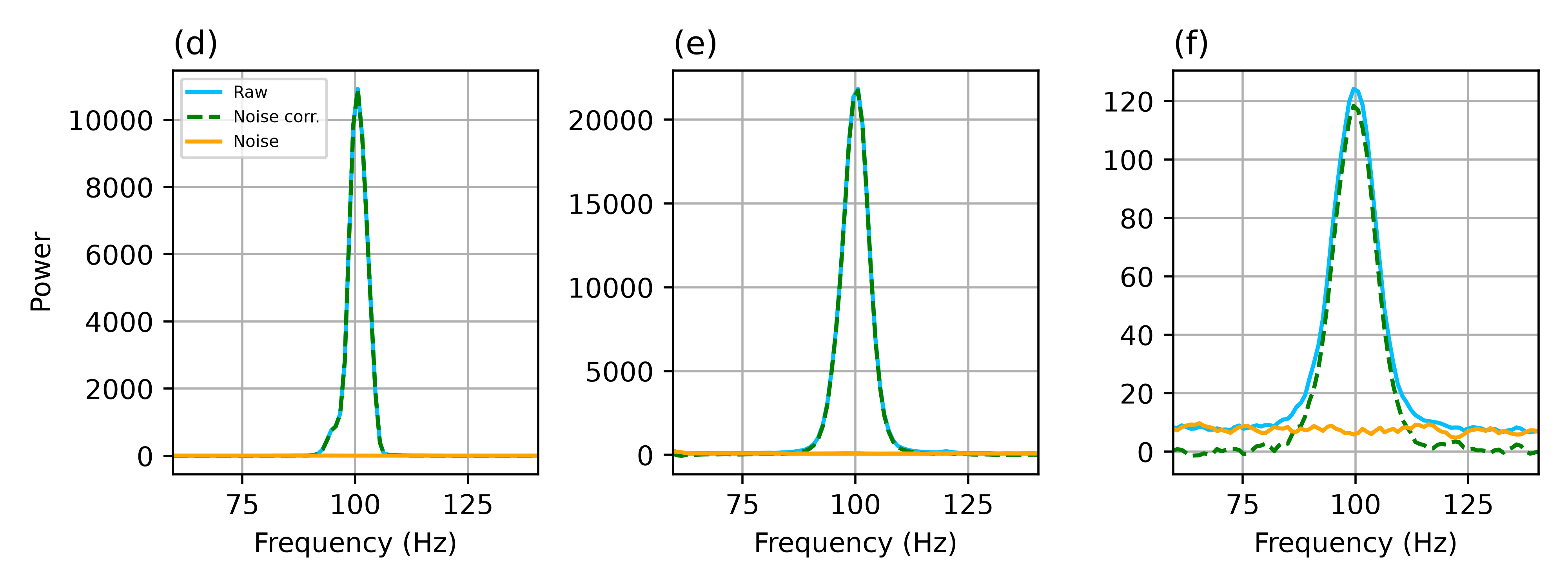}
    \includegraphics[width=\linewidth]{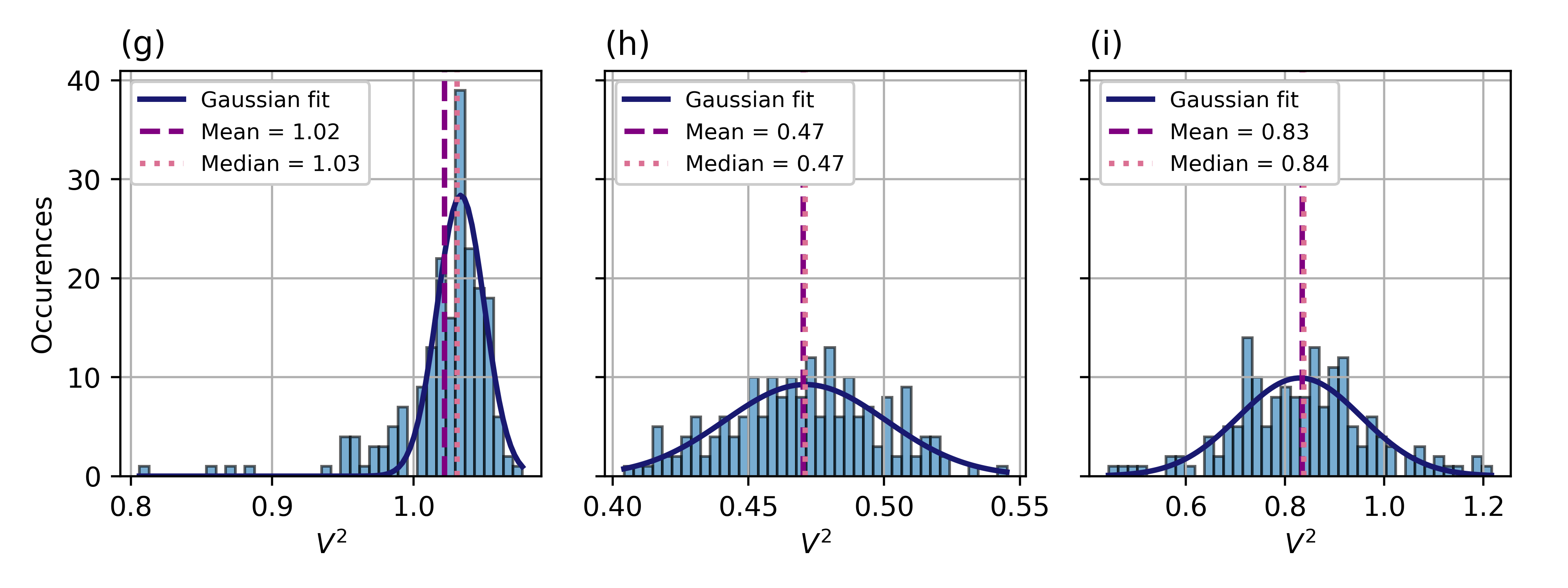}
    \caption{(a-c): Evolution of raw intensity fluxes measured with (a, d, g) the internal STS source, (b, e, h) a bright star (HD~172167), and (c, f, i) a faint star (HD~210418) in July 2025. The interferometric BC outputs are in cyan (I$_1$) and dark blue (I$_2$) lines, and the photometric BC outputs in purple (P$_A$), and in orange (P$_B$). (d-f): The estimated power spectral densities of the output I$_1$ are plotted as a function of the Fourier frequency. The raw power averaged over all scans is in cyan, the estimated noise power is in orange, and the average power over all scans corrected by the noise is in dashed green. (g-i): Histograms of the squared visibilities with a Gaussian fit to visualise the distribution. Note that the estimated squared visibilities obtained from the power spectra (panels d-i) do not precisely match the contrast of the raw fringes (panels a-c). This is due to 1. a low sampling rate of the raw fringe data, which underestimates the fringe contrast and 2. a bias that may be introduced by errors in the estimation of the effective spectral bandwidth of CHARIOT.}
    \label{fig:dataprocessingsteps}
\end{figure*}

After a few adjustments, the former JouFLU control software has been repurposed to acquire data from CHARIOT. 
The JouFLU software itself is based on the CLASSIC control software. 
The acquired data are the four values of flux of four pre-chosen pixels, where the BC outputs are imaged.
To measure stable temporal interference fringes, the camera is synchronised with the motorised translation scanning stage (Stage$_{scan}$) that controls the OPD$_{scan}$. The latter's position is dithered with a triangular waveform of peak-to-peak amplitude of $\pm 27.5\,\upmu$m, and of half a period of $511\,$ms, which scans the OPD$_{scan}$ between $-55\,\upmu$m and $+55\,\upmu$m (during that half period), with a speed of $215\,\upmu$m/s (details in ~\cite{Scott:2013}). The camera acquires frames at a rate of 500\,fps, which gives an interferogram of 250\,samples, constituting what is called a scan. 
Figure~\ref{fig:dataprocessingsteps}~(a-c) shows such scans of the four BC outputs (interferometric in cyan and dark blue, photometric in purple and orange) measured on the STS internal source (a, d, g), the bright star Vega HD~172167 (b, e, h) and the fainter star HD~210418 (c, f, i).
Thus, the resulting fringe frequency is $100\,$Hz, which has the advantage of minimizing the atmospheric piston perturbation. Figure~\ref{fig:dataprocessingsteps}~(d-f) shows the fringe power at $100\,$Hz estimated from the I$_1$ interferograms of the top row.
The power spectral density is shown for the average over all scans in cyan, for the noise level in orange, and after noise subtraction in dashed green (more details in Section~\ref{sec:pipeline}).
Finally, Fig.~\ref{fig:dataprocessingsteps}~(g-i) presents the histograms of the squared visibilities, from which the mean or median can be calculated, with a Gaussian fit to visualise the distribution.

Throughout the observations in 2024, the camera image was binned to increase the flux captured per BC output, aiming for a better signal-to-noise ratio (SNR). However, a more detailed study of the binning method carried out with the STS source in July 2025 concludes that this is not a favourable method. For bin sizes different than 1, the splitting ratio cannot be conserved for lower STS fluxes and detection of faint signals is reduced. Appendix~\ref{sec:binning} provides more details on the binning algorithm, the proportion of captured flux, as well as the results of the comparison study. Thus, the binning was applied with a bin size of 1 by default during the run of July 2025, even if a significant amount of light is lost (Fig.~\ref{fig:binning}).

Before the acquisition of on-sky interferometric data, several optimisation processes are performed. First, a raster scan of the motorised mirrors ZM allows for optimising the amount of flux injected into the BC. Then, the fringe packet is searched by moving the delay line cart around the position predicted by the baseline solution model (on a mm scale offset) because (1) the optical paths are different when observing on-sky as opposed to on the STS and (2) a normal zero OPD drift of the observatory baseline carts is observed across data sets (on the order of several hundreds of micrometers per night), as limited by the current baseline solution. Finally, the data set is composed of a shutter sequence which is as follows and is saved in a single .fit file:
\begin{enumerate}
    \item Background (25 scans): both shutters are closed
    \item First input beam (25 scans): one shutter is opened while the other one is closed
    \item Second input beam (25 scans): the opposite shutter configuration as the previous step is set
    \item Fringes (200 scans, but low SNR data can be skipped automatically): both shutters are opened
    \item Additional final sequence step (200 scans): either off-fringe (both shutters open, but moving the cart out of the coherence length by $\sim 1\,$mm) or more background data (same as the first step).
\end{enumerate}

%%%%%%%%%%%%%%%%%%%%%%%%%%%%%%%%%%%
\subsection{Data analysis methods}
\label{sec:pipeline}
The data analysis for this work is based on the Redfluor analysis pipeline from the CHARA Array, which has been developed in C for the analysis of CLASSIC and CLIMB data, as well as FLUOR and JouFLU data (see the CHARA website and technical reports, e.g. \cite{tr96} for more details and \cite{Bouquin:2004} for the core calculation). 
As part of the Redfluor pipeline, each interferometric scan is photometrically calibrated by the photometric outputs, and the power spectrum is calculated. Subsequently, the power spectrum of the noise is estimated from the off-fringe sequence (see point (v) of Sec.~\ref{sec:daq} and \cite{tr96}). The resulting average power spectrum of the noise is then subtracted from all interferometric power spectra. After a noise-rejection process that eliminates unsuitable noisy scans, the squared visibility $V^{2}$ is estimated by integrating the noise-corrected fringe power S (over a window determined by fitting a Gaussian function on the average of all interferometric power spectra located around $100 \,$Hz) as follows: 
\begin{equation}
    V^2 = 4 \Delta k v S.
    \label{eq:V2}
\end{equation}
With the bandwidth $\Delta k$ (in units of wavenumber) and the scanning speed of the OPD (here: OPD$_{scan}$) $v$. Note that all mentions of the squared visibilities refer to squared visibilities that are not calibrated by calibrator star squared visibilities.

In case of low SNR signal, the simultaneous photometric calibration would propagate noise to the interferometric signal and to the squared visibility estimation, because of the uneven (25:75) splitting ratio between the photometric and the interferometric channels. As a consequence, when the measured flux of the photometric channels has a low SNR, we may decide to command the data reduction pipeline to ignore the simultaneous photometric calibration, and the data measured during the first three steps of the shutter sequence are used as a photometric reference (which has a higher SNR).

In the output of the pipeline, the mean and median over all raw squared visibilities estimated from all scans in the data set are given, as well as their standard deviation (estimated from the median absolute deviation for the latter). The pipeline computes those products via two different methods. The first was written for the two-beam combiner CLASSIC data (DIRECT~PS) which do not have photometry channels, and the second was written for the FLUOR data (FLUOR~PS) which have photometry channels (allowing for simultaneous photometric calibration). Hence, we show the mean squared visibility results from the DIRECT~PS method ($\bar{V}^2_{DPS}$ or \lstinline{V2_SCANS_DIR} in the pipeline) for the May 2024 data sets and from the FLUOR~PS method for all later data sets.

Following the conclusions of \cite{Nunez:2017a}, the choice of the statistical calculation for squared visibility estimation can influence the systematic bias and precision of the resulting value. In this paper, we choose to present the median of the squared visibility $\bar{V}^2_{med}$ (\lstinline{V2_MEDIAN} in the pipeline) obtained from the FLUOR~PS calculation method because it is more resilient to outliers. The results can differ depending on the method, especially due to different noise subtraction methods and to a data rejection only applied with FLUOR~PS. In the following, the notation $\bar{V}^2$ describes the squared visibility estimator in general, encompassing the different methods and their mean and median. The subscript denotes the specific method.

Here, we are also interested in the precision, which we calculate as the standard error of the mean ($\sigma_{\upmu}$), i.e. the standard deviation $\sigma$ (the median absolute deviation) of the mean (median) squared visibility divided by the square root of the number of scans $N$:

\begin{equation}
    \text{$\sigma_{\upmu}$} = \frac{\sigma}{\sqrt{N}}.
\end{equation}
Then, the relative standard error (RSE) is calculated in percent relative to the mean (or the median) of the squared visibility $\bar{V}^2$: 
\begin{equation}
    \text{RSE} = \frac{\text{$\sigma_{\upmu}$}}{\bar{V}^2}\cdot100\%.
\end{equation}

We use meta-analysis tools and concepts to assess repeatability and consistency or to obtain the overall mean of the squared visibility estimates over several data sets. The statistical methods of the analysis of variance (ANOVA) are applied, which helps to compare variations within a data set ("within study", i.e. within the 200 scans) with variations between the data sets ("between study"); see, e.g. \cite{quene:2024, Veroniki:2016, rice:2006}. For several data sets of the same target, the combined mean $\bar{V}_{c}^{2}$ of the squared visibility, the pooled standard deviation $\sigma_{pool}$, and the corresponding $\sigma_{\upmu, pool}$ can be calculated as follows:
\begin{equation}
    \bar{V}_c^{2} = \frac{\sum_i^k (N_i \cdot \bar{V}_i^2)}{N_{Total}},
\end{equation}
\begin{equation}
    \sigma_{pool} = \sqrt{\frac{\sum_i^k(N_i-1)\cdot\sigma_i^2 + \sum_i^k N_i\cdot(\bar{V}_i^2-\bar{V}_c^{2})^2}{N_{Total}-1}},
    \label{eq:spool}
\end{equation}
\begin{equation}
    \sigma_{\upmu, pool} = \frac{\sigma_{pool}}{\sqrt{N_{Total}}}.
\end{equation}

where $N_{Total} = \sum_i^k N_i$, with $i$ being the estimate index (each consisting of multiple scans), and $k$ the number of estimates (number of data sets included in the meta-analysis). The numerator of Eq.~\ref{eq:spool} consists of the sum of the squared deviations (sums of squares) for the squared visibilities of the within and between studies. The between-study standard deviation $\tau$ is calculated using the DerSimonian and Laird method: 
\begin{equation}
   \tau = \sqrt{\max \Biggl\{0, \frac{\sum_i^k w_i \cdot (\bar{V}_i^2-\bar{V}_c^2)^2 - (k-1)}{\sum_i^k w_i- \frac{\sum_i^kw_i^2}{\sum_i^k w_i}} \Biggr\}}, 
    \label{eq:tau}
\end{equation}

where $w_i = \sigma_i^2/N_i$ are the weights of each data set. With $\tau$, we calculate the between-study coefficient of variation $CoV_\tau$ of the combined mean:
\begin{equation}
    CoV_\tau = \frac{\tau}{\bar{V}_c^{2}}
\end{equation}

For CHARIOT, we chose to take into account time-dependent and systematic effects, such as drifts, thus assuming a random-effects model rather than a fixed-effects model \citep{borenstein:2010}. This influences the definitions of $\sigma_{\upmu, pool}$ and $\tau$.

%%%%%%%%%%%%%%%%%%%%%%%%%%%%%%%%%%%%%%%%%%%%%%%%%%%%%%%%%%%%%%%%%%%%%%
\section{Results of CHARIOT at the CHARA Array}

%%%%%%%%%%%%%%%%%%%%%%%%%%%%%%%%%%%
\subsection{Laboratory characterisation}
\label{sec:stscharac}
The STS light is injected into CHARIOT to conduct its laboratory characterisation. The internal squared visibility of the instrument has been estimated on several days to assess the stability and the reproducibility of the measurements. For five measurements taken on different days in July 2025 with BC-25, a combined mean of $\bar{V}^{2}_{c, med} = 1.028$ is calculated ($\sigma_{\upmu, pool}$ = 0.001), with a pooled standard deviation $\sigma_{pool} = 0.029$, N$_{total} = 950$, and a between-study standard deviation $\tau =  0.022$. The variation between the values is $CoV_\tau = 2.2\%$. Variations on the few-percent level for squared visibility measurements on different days have also been observed for other instruments such as MYSTIC at CHARA (Anugu~et~al., priv. comm.). For more details, Table~\ref{tab:STSvis} in Appendix~\ref{sec:appendixsts} shows the extracted squared visibility $\bar{V}^2$ with its uncertainties for several measurements on different days with BC-22 and BC-25 recorded with the maximum STS flux. For comparison, the results for $\bar{V}^2$ measured with BC-22 from May 2024 and October 2024, which are presented in the first two rows of Table~\ref{tab:STSvis}, are around $\bar{V}^2=0.8$, with an RSE of less than 1\%. 
We underline that an image binning with bin sizes different from 1 underestimates the extracted squared visibility compared to binning with a bin size of 1. Finally, for both BC-22 and BC-25 beam combiners, the instrumental CHARIOT squared visibilities measured at the CHARA array are higher than the value $V=0.82$ ($V^2 = 0.67$) determined in an earlier characterisation phase of the project \citep{Siliprandi:2024}. As the instrumental squared visibility is measured with the STS inside the CHARA laboratory, it corresponds to the intrinsic visibility of CHARIOT, without taking into account effects from the atmosphere or the beam path from the CHARA telescopes. 

\begin{figure*}
    \centering
    \includegraphics[width=\linewidth]{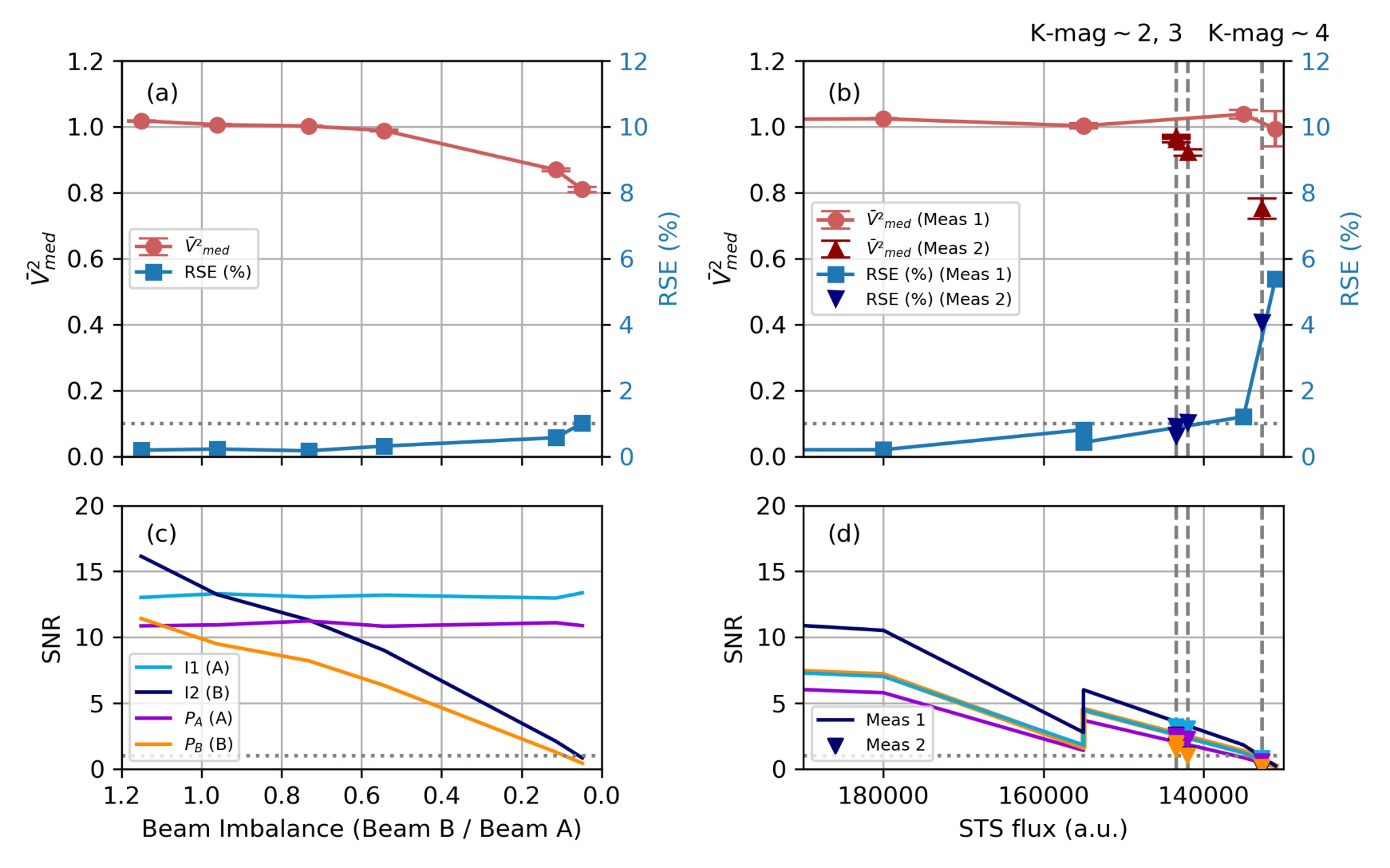}
    \caption{Laboratory characterisation of CHARIOT's sensitivity to flux reduction when reducing the power in only one beam (a) and when reducing the power in both beams (b) with the STS source. The estimated squared visibility $\bar{V}^{2}_{med}$ with its $\sigma_{\upmu}$ as error bars are in red / dark red and the RSE is shown in blue / dark blue. In the left plot (a), both beams start with maximum STS flux, which corresponds to a beam imbalance of beam B / beam A = 1.15 (the STS flux differs between beams, see Section~\ref{sec:instru_description}). For this measurement, the flux in beam B is reduced. The right plot (b) shows the results from two different sets of measurements in which the STS flux is reduced while the signal stays roughly constant by increasing $G_{APD}$ (Meas~1), and in which the STS flux is set to simulate specific K-band magnitudes of 2, 3, and 4 located by vertical dashed lines (Meas~2). The horizontal dotted lines mark a RSE of 1\%. The bottom panels (c) and (d) show the corresponding SNR, with the horizontal dotted line indicating a value of~1. The SNR is calculated from the data during the shutter sequence when each beam is injected sequentially. For I$_1$ and I$_2$, which behave very similarly, the SNR for only one input are shown as examples (beam A and beam B, respectively). The colour-code shown in the legend in (c) also applies to (d). In plot (d), the solid line is data for Meas~1, while the triangles denote the data for Meas~2.}
    \label{fig:stsvistest}
\end{figure*}

In addition to the characterisation at maximum STS flux, two more tests were performed with BC-25 to simulate stellar sources and identify potential limitations. The first test consists in measuring the effect of reducing the flux in one arm by slightly decoupling one of the light beams (beam B) with the motorised coupling mirror ZM, until it reaches an imbalance (beam~B/beam~A) of 0.05, i.e. beam B has been reduced to 5\% of the flux in beam A (Fig.~\ref{fig:stsvistest}~(a, c)). The second one assesses the precision of the squared visibility estimation as a function of the target magnitude by gradually reducing the STS flux simultaneously in both arms (Fig.~\ref{fig:stsvistest}~(b, d)). For both tests, the SNR is calculated from data recorded during the shutter sequence when only one beam is injected (on the outputs I$_1$ and P$_A$, when input beam A is injected, and on the outputs I$_2$ and P$_B$, when input beam B is injected). Only one interferometry output is presented per input beam as I$_1$ and I$_2$ behave similarly.

Figure~\ref{fig:stsvistest} shows the estimated squared visibility $\bar{V}^2_{med}$ (red circles) and its RSE (blue squares) for these tests. When the power is reduced in only one beam (beam B), the squared visibility deviates significantly from 1 for an imbalance (beam~B/beam~A) stronger than 0.1, see Fig.~\ref{fig:stsvistest}~(a). At these points, the SNR for I$_2$ and P$_B$ have dropped to SNR < 3 for an imbalance of 0.1 and SNR < 1 for an imbalance of 0.05, see Fig.~\ref{fig:stsvistest}~(c). The RSE only increases slightly and just reaches the 1\% mark for an imbalance of 0.05.
Figure~\ref{fig:stsvistest}~(b, d) shows the results of the second test. Only the lowest flux values are displayed, because the squared visibility and its precision remain fairly constant for the rest of the high fluxes. It should be noted that the flux axis is in decreasing order, is in arbitrary units (a.u.), and is not linear, because it is internally set by moving a mirror that decouples light. In a first set of measurements (Meas~1), the STS flux is reduced while the gain is increased, so the measured signal stays fairly constant to simulate observation of fainter targets during which we would increase the gain of the camera. As a consequence, several points of different gains overlap on the graph at the same flux. When the gain is adjusted, the SNR also changes. Then, in a second set of measurements (Meas~2) the STS flux and $G_{APD}$ are adjusted so that the current measured signal matches the measured signal during previous observations of stars with known K-band magnitudes. Three of those measurements are indicated by vertical dashed lines for K-mag $\sim$(2, 3, 4). Their squared visibilities (dark red upward triangles) decrease with larger magnitude (lower SNR), while their RSE (dark blue downward triangles) increase. For K-mag\,$\sim$2, the RSE is still lower than 1\%, reaching the 1\% mark for K-mag\,$\sim$3 and is on the order of a few percent for a K-mag of 4.3. For all data in Meas~2, the SNR is fairly low: the SNR is $\leq3$ for K-mag\,$\sim$2, reaches values of 1 for K-mag\,$\sim$3, and drops to < 1 for K-mag$\,\sim$4. For more details on the measurements presented in this figure, see Table~\ref{tab:STSvistest}.

Based on these measurements as well as on simulated mock data with different noise levels, we observe that the squared visibility estimator diverges from the expected visibility, losing accuracy and precision, when one of the interferometric arms has a SNR value below a typical threshold of $\sim$3-6. This emphasises the importance of improving the SNR of the flux measured by CHARIOT to obtain accurate sub-percentage precision measurement for stars fainter than K-mag\,$\sim$2, as atmospheric turbulence may decrease the amount of flux coupled into the fibres.

%%%%%%%%%%%%%%%%%%%%%%%%%%%%%%%%%%%
\subsection{On-Sky qualification}
\label{sec:onsky}
This section presents three on-sky qualification steps of CHARIOT during three observing runs. The first one took place on the 2nd of May 2024 (UT) for the first light of the instrument with the BC-22 using the shortest baseline ($34.07\,$m) with the S1 and S2 telescopes. The second one took place on the 22nd of October 2024 (UT), using the same baseline S1-S2, following the back end optics system upgrade, which allowed for the simultaneous measurement of the interferometric signal and the photometric channels. Finally, the third one took place on the 26th, 27th and 28th of July 2025 (UT) after the integration of the BC-25 and using the S1-S2 and E1-E2 ($65.88\,$m) baselines to assess the instrument stability and precision.

An overview of the extracted squared visibilities of different stellar targets during three separate engineering runs is shown in Fig.~\ref{fig:visvsdate} and in the Appendix~\ref{sec:appendixvis}. Figure~\ref{fig:visvsdate} also shows their expected squared visibilities (black crosses), which are modelled from a uniform disk in ASPRO with stellar diameters taken from SearchCal, and without accounting for the CHARIOT internal loss of visibility, the internal loss from the optical path of CHARA upstream to CHARIOT, which includes the polarisation mismatches, the chromatic dispersion, and the residual atmospheric effects, which can overall account for $\sim$15\% in squared visibility loss~\citep{Anugu:2026}.

The squared visibilities reported hereafter are all photometrically calibrated. However, the visibility losses in the transfer function resulting from the resolved stellar diameter are not calibrated out. 
All seeing estimates were acquired from the wave-front sensors of the telescope AO systems.

\begin{figure*}
    \centering
    \includegraphics[width=\linewidth]{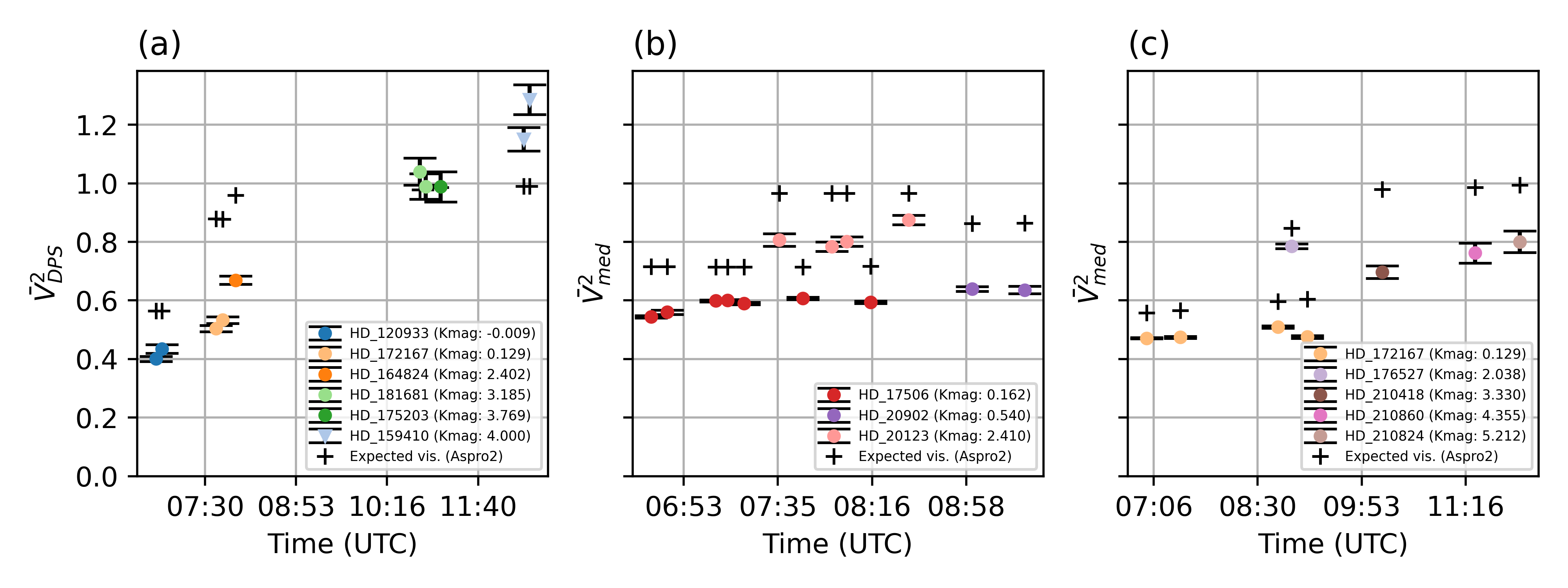}
    \caption{Overview of the extracted squared visibilities for different observation nights during three separate engineering runs: (a) May 2024 with BC-22, (b) October 2024 with BC-22, and (c) July 2025 with BC-25. The circle markers are the visibilities estimated from the difference signal of the two interferometric outputs and the triangle markers is from output I$_1$, the error bars are given by $\sigma_{\upmu}$ and the black crosses are the expected instrumental squared visibility values taking into account with ASPRO the finite stellar diameter modelled as a uniform disk. 
    All squared visibilities are photometrically calibrated, whereas the coherence losses due to the resolved stellar disk are not calibrated out. The intrinsic visibility loss from the CHARA array is also not included in the ASPRO model. }
    \label{fig:visvsdate}
\end{figure*}

%%%%%%%%%%%%%%%%%
\subsubsection{First light of CHARIOT in May 2024}

During this initial engineering run for CHARIOT in May 2024, the objectives were to obtain the first stellar light, to identify and address technical issues, and to understand the instrument's performance and limitations. Hence, the targets were selected as a sample of different K-band magnitudes. The seeing varied following values of r$_0$ from $\sim$6 to $\sim$16\,cm (increasing towards the end of the night).

As the instrument was still in a preliminary state of being set-up (e.g. two-lens system with higher losses due to aberrations, no photometric taps, BC-22, camera vibrations not subtracted, LNPs not fully optimised, and control software still being adjusted to the new hardware), the median squared visibility $\bar{V}^{2}_{med}$ could not be reliably extracted; thus only the DIRECT~PS squared visibility $\bar{V}^{2}_{DPS}$ is presented in Fig.~\ref{fig:visvsdate}~(a). For more details, Table~\ref{tab:targets2024A} in the Appendix lists the observed targets, the seeing conditions, and the extracted squared visibilities of CHARIOT in May 2024.

Despite these limitations, the CHARIOT instrument was able to observe six different targets, ranging in K-band magnitudes from 0 to 4 for which squared visibilities were estimated with an RSE of a few percent. The values of $\bar{V}^{2}_{DPS}$ range from 0.4 to $\sim$0.7 for the brighter targets observed during the first part of the night. 
Around the middle of the observation night, we experienced technical issues that included malfunctioning of the scanning stage controller and signal loss, which is visible as a gap in Fig.~\ref{fig:visvsdate}~(a). While we were able to recover the signal enough to continue observations --with SNR<1--, measurements after the gap have larger squared visibility values ($\bar{V}^{2}_{DPS}$ $\sim$1.0-1.3) and higher uncertainties than before. We cannot say with certainty whether this is solely due to the lower signal strength of the fainter targets or the effect of residual technical issues. For the faintest target (HD~159410), we suspect that the frequency pollution of the camera vibration close to the fringe frequency artificially increases the amplitude of the power spectrum, from which the squared visibility is calculated. Moreover, only a single interferometric output (I$_1$) is used for the data analysis, because we centred it on the output optical axis to increase its signal strength, thus removing the other interferometric output from the limited field of view. For the other targets, the difference signal of the two interferometric outputs is used, which seemingly reduces the effect of the camera vibration.
Therefore, we believe that those values do not exhibit any astrophysical phenomenon. 
As hard- and software adjustments to the instrument took place throughout the night, consistent results are not expected. However, measurements of the same target in a short period of time (on the order of tens of minutes) result in consistent squared visibilities.

%%%%%%%%%%%%%%%%%
\subsubsection{Instrumental squared visibility and stability - October 2024}

The second engineering run in October 2024 followed the integration of the new imaging optics system solution, which was the major upgrade. Throughout the observation nights, adjustments and repairs were required, which affected the measurements. This is partially an effect of re-activating the old JouFLU hardware and upgrading it to CHARIOT, which was in need of maintenance after many years of inactivity. Figure~\ref{fig:visvsdate}~(b) shows the squared visibilities $\bar{V}^{2}_{med}$, and their $\sigma_{\upmu}$ as error bars estimated on the three stars HD~17506, HD~20902, and HD~20123 observed on the 22nd of October 2024 (UT). Their K-band magnitudes range from 0.16 to 2.41 and the seeing conditions were moderate (around $5 - 7\,$cm). A detailed list of the squared visibilities using the two extraction methods, $\bar{V}^{2}_{med}$ and $\bar{V}^{2}_{DPS}$, can be found in Table~\ref{tab:targets2024B} of the Appendix~\ref{sec:appendixvis}.

We measured squared visibilities with RSEs of less than 1\% to a few percent on the brightest star HD~17506, which is comparable to similarly bright stars observed in May 2024, and shows the good interferometric performance reachable with CHARIOT. However, we could not demonstrate a better limiting magnitude than during the first run because of faulty behaviour of several hardware components, which resulted in the loss of flux. Note that the simultaneous photometric calibration has been ignored for the three last points of this target.

For seven measurements of HD~17506 (K-mag = 0.162), the combined mean of the squared visibilities is calculated as $\bar{V}^{2}_{c, med} = 0.585$ ($\sigma_{\upmu, pool}$ = 0.002), with a pooled standard deviation of $\sigma_{pool} = 0.061$, $N_{total} = 1283$, and a between-study standard deviation $\tau = 0.021$. For two measurements of HD~20902 (K-mag = 0.52), the combined mean of the squared visibilities is $\bar{V}^{2}_{c,med} = 0.638$ ($\sigma_{\upmu, pool}$ = 0.007), with $\sigma_{pool} = 0.122$ and $N_{total} = 294$. With a between-study standard deviation $\tau = 0$, the variation between measurements is solely due to the sampling error. Finally, for the faintest target in this run, HD~20123 (K-mag = 2.41), four measurements were obtained within a time span of around one hour. Here, the combined mean of the measured squared visibilities is $\bar{V}^{2}_{c,med} = 0.815$ ($\sigma_{\upmu, pool}$ = 0.009), with $\sigma_{pool} = 0.225$ and $N_{total} = 680$. The between-study standard deviation is $\tau =0.039$. By compiling those results, over $\sim 3\,$h, we thus show that the instrument is fairly stable with $CoV_\tau = 3.6\%$, $0\%$ (with $\tau = 0$), and $4.7\%$, respectively, for the three aforementioned targets; the corresponding precisions of the combined mean for each target are, respectively, $0.29\,\%$, $1.11\,\%$, and $1.06\,\%$ (RSE calculated from $\sigma_{\upmu, pool}$ and $\bar{V}^{2}_{c,med}$).  

The expected squared visibilities (dark crosses) are calculated using a uniform-disk model using JMMC's ASPRO \citep{ASPRO2:2016}, with diameters of $5.044\,$mas, $3.265\,$mas and $1.586\,$mas, respectively, for HD~17506, HD~20902 and HD~20123. Over the span of the measurements, the expected values are on average $V^{2}_{exp} = 0.717$, $V^{2}_{exp} = 0.86$ and $V^{2}_{exp} = 0.966$, respectively. For the three targets listed above, the instrumental squared visibilities (also known as transfer function) are calculated from the combined mean as $\bar{V}^{2}_{c,inst} = $ $ 0.816 \pm 0.002$, $0.742 \pm 0.008$, and $0.843 \pm 0.009$, respectively, where the uncertainties are calculated as $\sigma_{\upmu, pool}/V^{2}_{exp}$. Hence, the total CHARIOT + CHARA instrumental squared visibility is estimated to $\sim$0.80 $\pm$ 0.04 (the uncertainty is the standard deviation of the three values). For comparison, FLUOR had a transfer function of $\sim$85\,\% \citep{Absil:2006}.

%%%%%%%%%%%%%%%%%
\subsubsection{Instrument sensitivity and precision - July 2025}

Figure~\ref{fig:visvsdate}~(c) shows a selected subset of squared visibilities $\bar{V}^{2}_{med}$, from one of the three nights of the July 2025 observing run, the 27th of July 2025 (UT). The selection of the targets presented here is chosen to match their magnitudes to the measured fluxes of the STS laboratory characterisation demonstrated in Table~\ref{tab:STSvistest}. Table~\ref{tab:targets2025A} in the Appendix provides more details on the targets, the seeing conditions, the extracted squared visibilities $\bar{V}^{2}_{med}$ and $\bar{V}^{2}_{DPS}$ and their precisions. It is to be noted that no technical issues hindered the observation nights, which seem to have all been fixed during the preparation as well as during previous engineering runs. This allowed for quick and steady measurements under comparably consistent conditions.

Despite the worse seeing conditions (r$_0$ around $1-6\,$cm), the precisions on the squared visibility estimation are better than during previous engineering runs. Sub-percent precision is achieved for the brightest target (HD~172167). Compared to observations from October 2024, the RSE of $\bar{V}^2_{med}$ improved by a factor of $\sim$2.2 for K-mag~$\sim$2. This can be attributed to the improved throughput obtained with BC-25, thus increasing the data SNR. For the July 2025 measurement of HD~176527 presented here, the SNR is between 1 and 3 for all outputs. This is a factor $\sim$10 higher than the lowest SNR of the HD~20123 measurements from October 2024. When comparing the on-sky measurement precisions in Table~\ref{tab:targets2025A} with the laboratory sensitivity characterisation using the STS in Tables~\ref{tab:STSvis} and~\ref{tab:STSvistest}, a similar trend with similar precision values can be observed: bright flux can achieve sub-percentage precision, while the faintest flux has precisions of around 4-5\%. This indicates that with these experimental conditions, the atmosphere is not currently limiting individual measurements. Thus, pushing for better precision for fainter stars requires improvements by fine-tuning experimental parameters --especially those that increase the SNR--, such as flux extraction method, integration time (within the limitations of atmospheric effects), number of scans, or background noise. 

Moreover, the limiting magnitude was improved compared to previous engineering runs. Fringes were measured for a star of a K-band magnitude of 5.212 (HD~210824) -- gaining an order of magnitude in sensitivity, despite the poorer seeing. The measured squared visibility is $\bar{V}^{2}_{med} = 0.80 \pm 0.04$, with an RSE of 4.61\%, which is promising for future science observations.

During the night, the stellar target HD~172167 (Vega, K-mag = 0.129) was observed four times within a time window of around two hours (with two different gains and varying seeing conditions). Using the respective measured and expected squared visibilities of the four individual measurements, the instrumental squared visibilities are calculated and statistically assessed. The combined mean is $\bar{V}^{2}_{c,inst} = 0.821$ ($\sigma_{\upmu, pool} = 0.003$), with $\sigma_{pool} = 0.072$, $N_{total} = 637$, and a between-study standard deviation of $\tau = 0.028$. 
While we observe a difference between the measured and the theoretical squared visibilities, resulting mostly from coherence losses not fully accounted for in our ASPRO modelling (see Section~\ref{sec:onsky}), we also see that the transfer function is stable within $CoV_\tau = 3.4\%$.

%%%%%%%%%%%%%%%%%%%%%%%%%%%%%%%%%%%%%%%%%%%%%%%%%%%%%%%%%%%%%%%%%%%%%%
\section{Discussion and Conclusions}

Over several years, the former JouFLU instrument has been taken apart, adjusted, and upgraded, thus becoming CHARIOT. We have reported on the instrument description, the laboratory calibration, and the first on-sky fringes with their raw squared visibilities and precisions of the CHARIOT instrument at the CHARA Array. Throughout and after each engineering run of CHARIOT, subsystems such as the output optics system have been updated to improve the functionality and/or performance. In addition to a change in beam path and optical components, the main differences consist in the C-RED One camera, which replaces the former NICMOS camera, and the photonic beam combiner chip, which replaces the degraded MONA fibre-based beam combiner. The two-telescope astrophotonic beam combiner of CHARIOT was fabricated using ultrafast-laser inscription in a low-OH Infrasil quartz glass with high transmission properties over the whole K-band.
The CHARIOT instrument is planned as a plug-and-play testbench, and for that purpose its beam combiner is fibre connectorised with commercial FC/PC connectors at the input. Swapping the beam combiner can be completed within a few hours, with some adjustments to the optics. With this ability, novel astrophotonic components can be rapidly tested on-sky at the CHARA Array. With fibres, the insertion loss of the beam combiner increases from $\sim$1.1\,dB (bare chip) to $\sim$3.9\,dB, resulting in a total BC throughput of $\sim$40\%. As a comparison, the  fibre-connectorised BC of GRAVITY has a mean throughput of 55\% in the K-band~\citep{Perraut:2018}. A simple way to improve throughput in CHARIOT would be to reduce the output fibre length of the BC (input fibres are phase matched), which is much longer than necessary ($\sim$2.5\,m), due to the ongoing instrument development. As presented in Table~\ref{tab:chariotinstr}, the transmission losses are $\sim$0.1\,dB/m at 2.15\,$\upmu$m and rises to 0.5\,dB/m at 2.3\,$\upmu$m. For instance, by reducing the fibre length by $\sim$2\,m, the transmission at the reference wavelength of 2.3\,$\upmu$m would increase from $\sim$63\% to $\sim$79\%.

CHARIOT has observed its first on-sky fringes in May 2024. The squared visibilities estimated during the observations of three engineering runs (May 2024, October 2024, and July 2025) are presented and compared. Here, the Redfluor pipeline from CHARA is used to calculate the squared visibility with two different methods, namely the median $\bar{V}^2_{med}$, and the mean $\bar{V}^2_{DPS}$. A general improvement of the measurement precision is observed, owing to the continuous instrument optimisation over the past years. CHARIOT has measured fringe squared visibility with sub-1\% RSE for stars with K-mag\,$\lesssim$\,2. The limiting magnitude for fringe detection in the K-band in poor to moderate seeing conditions is K-mag = 5.212 (target HD~210824). The on-sky precision limits are in good agreement with those obtained from laboratory tests, which qualifies CHARIOT's suitability as a testbench. As a consequence, we also demonstrate that the astrophotonic BC is approaching the best results of the former JouFLU instrument (sub-percent RSE for K-mag\,$\lesssim 3$). CHARIOT already detected fringes for fainter targets than JouFLU, which had a limiting magnitude K\,$\sim$4.5. With the AO system that has been added to CHARA after JouFLU decommissioning in addition to further improvements of the current setup, including data acquisition and processing, we are confident that CHARIOT can reach sub-1\% RSE for stars with K-mag up to 4, surpassing the performance of JouFLU. To the best of our knowledge, this is the first on-sky demonstration of a ULI-fabricated astrophotonic BC in the K-band.

As a metric to quantify the quality of the laboratory measurements and the on-sky fringes, the squared visibilities were extracted with their $\sigma_{\upmu}$ and RSE. For a meta-analysis, a comparison of different measurements of the same targets is performed, where the combined mean with $\sigma_{pool}$ and $\sigma_{\upmu, pool}$ is calculated, as well as the between-study standard deviation. The between-study variation $CoV_{\tau}$ for different measurements of $\bar{V}^2_{med}$ is on the order of a few percent. For VLTI/GRAVITY+, variations on the order of $\sim$10\% have been measured --albeit over much longer timescales--, and simulations and discussion of the causes of the variations can be found in \cite{SIMTERFERE}, with throughput variation after fibre coupling identified as the dominant source.

By calculating the ratio between the measured combined mean of a target with the expected squared visibility (from ASPRO), the instrumental squared visibility of CHARIOT + CHARA is estimated to $\sim$0.80 $\pm\,0.04$ (October 2024) and $0.821 \pm 0.003$ (July 2025). For comparison, the laboratory characterisation of \cite{Siliprandi:2024} reported a (non-squared) visibility of $\sim$0.82. 

There is, however, an indication of bias in the measurements of the squared visibility: firstly, the overall system bandwidth of CHARIOT requires an updated measurement to accurately determine the scaling factor for squared visibility estimators using the fringe power spectrum. Secondly, laboratory characterisation with the internal calibration source at different power levels shows that the accuracy of the measured squared visibilities decreases for low SNR data (SNR\,$\lesssim$\,6). Therefore, the low SNR data presented here may be subject to bias. This emphasises the importance of further increasing the sensitivity of CHARIOT.

For future optimisation of the CHARIOT instrument, there are several possibilities to improve the RSE and the instrument sensitivity. The MIRC-X instrument is also considered to be used in parallel as a fringe tracker to correct for atmospheric turbulence during measurement. This would allow for a longer integration time, which further improves the sensitivity of CHARIOT. In the long term, other subsystem updates are considered, such as the integration of two more paths in order to inject the light of four telescopes into CHARIOT and a cryo-cooled beam-combiner system, which reduces background noise, thus paving the way to nulling interferometry developments.

%%%%%%%%%%%%%%%%%%%%%%%%%%%%%%%%%%%%%%%%%%%%%%%%%%%%%%%%%%%%%%%%%%%%%%
\section*{Acknowledgements}

This work is based upon observations obtained with the Georgia State University Center for High Angular Resolution Astronomy Array at Mount Wilson Observatory. The CHARA Array is supported by the National Science Foundation under Grant No. AST-2034336 and AST-2407956. Institutional support has been provided from the GSU College of Arts and Sciences, Office of the Provost, and Office of the Vice President for Research and Economic Development.

The research leading to these results has received funding from the Deutsche Forschungsgemeinschaft (DFG), grant number 506421303 ("NAIR-APREXIS"), the Bundesministerium für Bildung und Forschung (BMBF), grant number 03Z22AI1 ("Strategic Investment"), the H2020 Future and Emerging Technologies (820365-PHoG), the Science and Technology Facilities Council (ST/V000403/1), the Engineering and Physical Sciences Research Council (EP/X03299X/1), and the European Union’s Horizon 2020 research and innovation programme under Grant Agreement 101004719 (ORP). A.V.M. acknowledges support through the Erasmus+ program, and J.S. acknowledges funding through the James Watt PhD Scholarship by Heriot-Watt University. 
A.N.D. would like to thank A.~Schlumbohm for helpful conversations on the topic of statistical methods. This research has made use of the Jean-Marie Mariotti Center \texttt{Aspro}\footnote{Available at http://www.jmmc.fr/aspro} and \texttt{SearchCal}\footnote{Available at https://www.jmmc.fr/searchcal} services, which involve the JSDC and JMDC catalogues.

The authors are grateful to the referee, Chris Haniff, whose useful comments have significantly improved the quality of the paper.

%%%%%%%%%%%%%%%%%%%%%%%%%%%%%%%%%%%%%%%%%%%%%%%%%%%%%%%%%%%%%%%%%%%%%%
\section*{Data Availability}

The data underlying this article will be shared on reasonable request to the corresponding author.
%https://academic.oup.com/pages/open-research/research-data#Data%20Availability%20Statements


\bibliographystyle{mnras}
\bibliography{chariotbib} % if your bibtex file is called example.bib

%%%%%%%%%%%%%%%%%%%%%%%%%%%%%%%%%%%%%%%%%%%%%%%%%%
%%%%%%%%%%%%%%%%% APPENDICES %%%%%%%%%%%%%%%%%%%%%
\appendix
\section{Additional Camera Information and Characterisation}
\label{sec:camnoise}
%Data from 20241019, Bkg and Dark.

\subsection{Camera Unit Conversion}
The unit of the camera images is ADU, which can be converted into electrons (e$^-$) using:
    \begin{equation}
        N_{e^-} = \frac{G_{Sys}}{G_{APD}} \cdot N_{ADU}, 
    \end{equation}
    where $G_{Sys} = 1.9 \, e^-/\text{ADU}$ is the system gain (provided by the manufacturer) and $G_{APD}$ is the APD-Gain (adjustable).

\subsection{Camera noise}
Following the manufacturer's provided measurement procedures, we measure a dark current (in Global Reset Single (GRS) readout mode) of around $1300 \,e^-\cdot s^{-1}$ with the cap closed on the camera at room temperature. In addition, the readout noise (RMS noise) is measured to less than $1 \,e^-$ (in CDS readout mode) with APD gains $\geq 40$ ($\sim 16 \,e^-$ for APD gain = 2).

To better understand the background (BKG) noise of the CHARIOT C-RED One camera, BKG images were recorded at camera settings comparable to those during observation nights (CDS readout mode, fps = $500 \,$Hz).  %Data from 2024-10-19
As the APD gain $G_{APD}$ may be adjusted for targets of different brightness during an observation night, BKG images were recorded for different $G_{APD}$ values, to asses their influence on the noise. The BKG images were analysed per pixel as a function of time as well as across the detector to obtain the BKG offset and noise. Temporal (t) statistics are taken over the number of scans (here: 2600), spatial (x,y) statistics are taken over the pixels of the detector (excluding the first row, which contains reference pixels). Table~\ref{tab:cred_noise} shows the BKG offset (2nd column), the BKG standard deviation (3rd column), and the non-uniformity (4th column) for $G_{APD}$ values between 2 and 100 (1st column). The results with and without the (warm) camera cap only differ by few ADUs, thus shown here are the results with the cap off. The offset is the spatial median over the detector of the temporal mean value of a BKG measurement: Med(Mean(Pixel-Offset)$_t$)$_{xy}$. The standard deviation (StDev, $\sigma_{bkg}$) is the BKG noise that we can expect during measurements at different gain values. It can provide the basis for a SNR estimation. Shown here is the spatial median of the temporal mean: Med(Mean(Pixel-$\sigma_{bkg}$)$_t$)$_{xy}$. The non-uniformity describes the variation between the pixels, measured as the StDev of the deviation of each pixel offset (temporal mean) from the detector (spatial) median: StDev(Dev(Mean(Pixel-Offset)$_t$)$_{xy}$). 
The BKG offset increases linearly with $G_{APD}$, thus the dark and BKG current shot noise depend on $G_{APD}$. 
The measured BKG noise has contributions from the dark and BKG current shot noise, the readout noise, as well as non-uniformity when more than one pixel is considered.

As confirmed in a separate measurement, the readout noise is independent of the $G_{APD}$ in units of ADU (it decreases with gain in units of e$^-$) per pixel. The non-uniformity increases with $G_{APD}$, which should be taken into consideration, e.g. when binning is applied or when the selected pixels are changed.
\begin{table}
    \centering
        \caption{C-RED One camera in CHARIOT: measured BKG offset, noise, and non-uniformity for different $G_{APD}$ values.}
    \label{tab:cred_noise}
    \begin{tabular}{cccc}
         \hline
         $G_{APD}$ & BKG Offset & BKG StDev & Non-uniformity \\         
                & (ADU) &  (ADU) & (ADU)\\         
         \hline
          2   & 192 & 18 & 16 \\
          10  & 202 & 19 & 21 \\
          20  & 213 & 21 & 89 \\
          50  & 244 & 34 & 109\\
          80  & 272 & 47 & 163\\
          100 & 294 & 57 & 201\\
         \hline
    \end{tabular}
\end{table}

\section{Additional Beam Combiner Characterisation - Splitting Ratio}
\label{sec:splitratio}
The splitting ratio of the beam combiner (Fig.~\ref{fig:setup}~(a)(i)) is measured for each path by injecting a single beam into the BC. 
Here, the STS is used at maximum flux and is coupled into each beam combiner input. 
The measured splitting ratios for tap and interferometric arms, see Table~\ref{tab:splitting_ratio}, are close to the design values of 25:75 (Tap:Interferometry) and 50:50 (I$_1$:I$_2$), see \cite{Benoit:2021, Siliprandi:2024}.
Eight individual splitting ratio measurements were carried out between the 23rd and 28th of July 2025, and the table shows the combined mean and pooled standard deviation.
The between-study standard deviation of the splitting ratio measurements lies between 0.002 and 0.004. 
For each measurement, a set of a few hundreds (typically 200) camera images is used. 
First, the flux of each BC output (I$_1$, I$_2$, P$_A$, P$_B$) is extracted from several background corrected pixels where the flux is imaged, then the splitting ratio is calculated and averaged.
The splitting ratio can also be extracted from the shutter sequence measurements as part of the JouFLU automated data acquisition. 
As single pixels are used in that case, this might lead to small differences compared to multi-pixel flux extraction. 
\begin{table}
    \centering
        \caption{Photonic beam combiner splitting ratios obtained with the STS calibration source. "Total" is the sum over all four outputs: I$_1$ + I$_2$ + P$_A$ + P$_B$. "Tap" is P$_A$ + P$_B$, which only considers the corresponding tap for the input, i.e. P$_A$ for input A and P$_B$ for input B, as the other output is zero. The ratio  (I$_1$ + I$_2$) / Total  is calculated as 1 - (Tap / Total). Given for each splitting ratio is the combined mean with the pooled standard deviations, which is calculated from eight measurements on different days. }
    \label{tab:splitting_ratio}
    \begin{tabular}{|l|c|c|c}
        \hline
                    & Input A & Input B & Design \\ \hline
        Tap / Total & $0.253 \pm 0.004$  & $0.265 \pm 0.006$ & $0.25$ \\ 
        (I$_1$ + I$_2$) / Total & $0.747 \pm 0.006$  & $0.736 \pm 0.008$ & $0.75$ \\ 
        I$_1$ / (I$_1$ + I$_2$) & $0.495 \pm 0.006$ & $0.505 \pm 0.007$ &$0.50$\\ 
        I$_2$ / (I$_1$ + I$_2$) & $0.505 \pm 0.006$ & $0.495 \pm 0.007$ &$0.50$\\ 
    \end{tabular}
\end{table}

\section{Flux extraction methods and image binning}
\label{sec:binning}
\begin{figure}
    \centering
    \includegraphics[width=\linewidth]{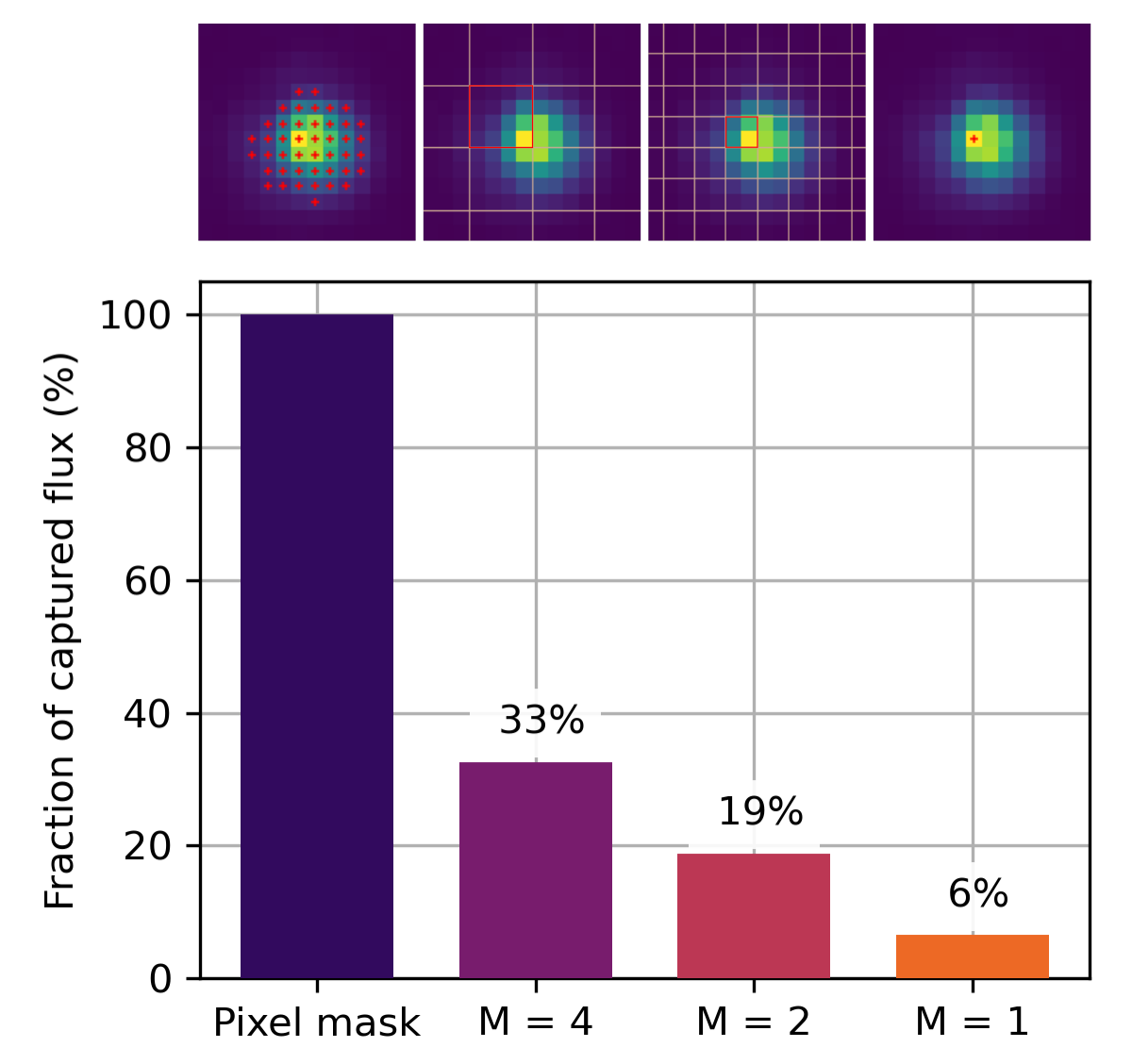}
    \caption{Fraction of the captured flux of output I$_1$ when the binning algorithm is applied to the C-RED One camera image with different parameters. The flux fraction is calculated relative to the pixel mask (left), which is assumed to collect 100\% of the beam. Here, the pixel mask contains 44 pixels, and the captured flux fraction is shown for the binning method with bin sizes M = 4, M = 2, and M = 1. The insets above the bars indicate the pixels from which the flux is collected (red crosses) for the pixel mask and the single pixel (M = 1) method. For the binning methods with M = 2 and 4, the bin borders are indicated (lines), with the bin that the maximum pixel coordinate falls in highlighted in red squares.}
    \label{fig:binning}
\end{figure}

With the installation of the C-RED One camera and the update of the output optics system, the PSFs of the waveguide outputs are not imaged on a single pixel, but instead spreads over a larger area on the detector. To estimate the extent of the beam, the STS is injected into one arm of the BC at a time, and a set of camera images of the outputs is analysed. For each output, the pixel coordinates with the maximum flux are identified. Then, a mask of pixels is created around them with values larger than the defined threshold $ADU_{th} = 3 \times \sigma_{bkg}$, resulting in a beam size on the order of 40~pixels. If the summed value of those pixels is assumed to represent 100\% flux, we can estimate the fraction of flux in the single pixel of maximum intensity: around 6\% of the total flux are captured in the single pixel at the beam's maximum, as shown in Fig.~\ref{fig:binning}.

As the existing JouFLU software is mostly used for data acquisition on CHARIOT, the data sent from the camera must match the format expected by the JouFLU software, which corresponds to one flux value (in ADU) per output. This requires any flux extraction method to be applied before the data are sent to the JouFLU software. Within the constraints of the existing software, a simple binning algorithm is included, which combines the flux from multiple pixels with the intention of increasing the SNR. The binning algorithm combines $M \times M$ pixels, with $M  \in \{1,2,4\}$ called the bin size. From the pixel coordinates $(x,y) = (0,0)$, the whole image is reshaped into a lower resolution image, where the ADU values of all pixels within a bin are summed. M = 1 corresponds to the original image, M = 2 is the sum of the value of 4 pixels, and M = 4 is the sum of the ADU value of 16 pixels, as represented in Fig.~\ref{fig:binning}.
After the installation of BC-25, the summed pixel values seemingly caused bit overflow for high flux values, resulting in an underestimation of the intensity, which distorted the data and required the implementation of an average of the binned pixel values rather than a sum. 
When binning is applied, the pixel coordinates that were selected in the unmodified image are converted to bin pixel coordinates by dividing them by the bin size M. 
The result is converted into an integer that truncates the quotient. 
Here, the location of the bins is not optimised: the maximum pixel can be at the edge of a bin, thus splitting the brightest part of the beam across bins (as we can see on the top part of the Fig.~\ref{fig:binning}, in which the selected bin pixel is higlighted in red).

In July 2025, a systematic study was conducted to evaluate the binning method regarding flux, splitting ratio, and visibility for different bin sizes (M = 1, 2, 4) and different levels of STS flux. 
Table~\ref{tab:binning} shows a subset of the measurements, from the highest to the lowest STS flux levels with different gains, and reveals that the detected flux actually reduces with the bin size $M$, and the BC's splitting ratio becomes brightness dependent. 
Also, the lowest STS intensity that could be detected with M = 1 could not be detected with M = 4, which defeats the goal of the binning algorithm. 
Therefore, the current binning algorithm is not a suitable method to improve the flux and observe low-brightness targets. 
From then on, M = 1 (i.e., no binning) has been applied to all the measurements in July 2025 - even if this means that only a small fraction (on the order of 6\%) of the overall flux is captured.

Any improvements to the algorithm would have to be made within the constraints of the existing software and data structures. Instead, future changes to the software are planned to enable complete images to be sent.

\begin{table*}
    \centering
        \caption{Evolution of the recorded counts per output and the calculated splitting ratio as well as the raw squared visibilities and their statistical uncertainties for binning applied with different bin sizes M, measured with the STS source on 23rd July 2025.}
    \label{tab:binning}
    \begin{tabular}{cccccccccccccc}
        \hline
        M & STS flux (a.u.) & G$_{APD}$ & P$_A$ & I$_2$ & I$_1$ & P$_B$ & Tap/Total & (I$_1$+I$_2$)/Total & I$_1$/(I$_1$+I$_2$) & $\bar{V}^{2}_{DPS}$ & $\sigma$ & $\sigma_{\upmu}$ & RSE (\%) \\ \hline
       1 & 370 & 5 & 525 & 2090 & 2050 & 900 & 25.6 & 74.3 & 49.5 &1.005 & 0.053 & 0.004 & 0.43 \\ %2025_07_23_NOSTAR_fld_001
      %  1 & 155 & 20 & 245 & 985 & 978 & 427 & 9.2 & 37.3 & 37.1 & 16.2 & & & &\\ %2025_07_23_NOSTAR_fld_002
      %  1 & 170 & 7 & 256 & 1070 & 1053 & 457 & 9.0 & 37.7 & 37.1 & 16.2 &  & & &\\ %2025_07_23_NOSTAR_fld_003
       1 & 135 & 100 & 237 & 984 & 955 & 410 & 25.0 & 74.9 & 49.2 & 0.976 & 0.062 & 0.005 & 0.52 \\ %2025_07_23_NOSTAR_fld_004
      %  1 & 130 & 100 & 37 & 165 & 156 & 62 & 8.8 & 39.2 & 37.1 & 14.7 & & & &\\ %2025_07_23_NOSTAR_fld_005
        1 & 126 & 100 & 14 & 80 & 78 & 32 & 22.5 & 77.4 & 49.3 & 0.75 & 0.88 & 0.06 & 8.08 \\ \hline %2025_07_23_NOSTAR_fld_006
        2 & 370 & 5 & 328 & 1341 & 1071 & 606 & 27.9 & 72.1 & 44.4 &   0.900 & 0.061 & 0.005 & 0.55 \\ %2025_07_23_NOSTAR_fld_012
      %  2 & 160 & 20 & 279 & 1129 & 914 & 515 & 9.8 & 39.7 & 32.2 & 18.1 &  & & &\\ %2025_07_23_NOSTAR_fld_013
      %  2 & 170 & 11 & 271 & 1091 & 892 & 494 & 9.8 & 39.7 & 32.4 & 17.9 &  & & & \\ %2025_07_23_NOSTAR_fld_014
        2 & 135 & 100 & 127 & 578 & 490 & 87 & 16.6 & 83.3 & 45.8 & 0.968 & 0.083 & 0.007 & 0.70  \\ %2025_07_23_NOSTAR_fld_015
      %  2 & 130 & 100 & 13 & 72 & 55 & 0 & 9.2 & 51.4 & 39.2 & 0 &  & & & \\ %2025_07_23_NOSTAR_fld_016
        2 & 126 & 100 & 4 & 32 & 25 & 0 & 6.6 & 93.3 & 43.8 & - & - & - & - \\ \hline %2025_07_23_NOSTAR_fld_017
        4 & 370 & 5 & 165 & 668 & 514 & 351 & 30.3 & 69.6 & 43.4 &0.806 & 0.021 & 0.002 & 0.22 \\ %2025_07_23_NOSTAR_fld_007
       % 4 & 172 & 20 & 273 & 1088 & 847 & 605 & 9.7 & 38.6 & 30.1 & 21.5 &  & & & \\ %2025_07_23_NOSTAR_fld_008
       % 4 & 170 & 20 & 253 & 1022 & 782 & 536 & 9.4 & 37.5 & 36.8 & 16.1 &  & & & \\ %2025_07_23_NOSTAR_fld_009
        4 & 135 & 100 & 65 & 285 & 209 & 117 & 26.9 & 73.1 & 42.3 & 0.899 & 0.055 & 0.004 & 0.50  \\ %2025_07_23_NOSTAR_fld_010
       % 4 & 130 & 100 & 4 & 30 & 14 & 6 & 7.4 & 55.5 & 25.9 & 11.1 &  & & & \\ %2025_07_23_NOSTAR_fld_011
        4 & 126 & 100 & 0 & 0 & 0 & 0 & - & -  & - & - & - & - & - \\ %no fringes
    \end{tabular}
\end{table*}

\section{squared visibility Tables}
This section includes detailed tables with the extracted squared visibilities for different measurements in the laboratory (Section~\ref{sec:appendixsts}) and on-sky (Section~\ref{sec:appendixvis}). Note the following differences between the experimental configurations for different measurements: BC-22, as well as pixel binning (M = 4), was used for the 2024 measurements, and BC-25 was installed without pixel-binning (M = 1) for the 2025 measurements. The squared visibilities are calculated for one interferometric output for the faintest target in May 2024. For all other targets in May 2024 and for all targets in October 2024 and July 2025, the difference signal of the two interferometric outputs is used for the calculation of the squared visibility. For May 2024, a two-lens system was used that reduced transmission and prevented using the photometric taps, and the polarization plates were not yet fully optimised, which likely lowered the contrast. The subtraction of a background pixel to remove the effect of camera vibrations in real-time was also not yet included in May 2024.

\subsection{Laboratory Characterisation}
\label{sec:appendixsts}

Table~\ref{tab:STSvis} shows the extracted squared visibilities with their respective uncertainties as reference measurements in the laboratory using the STS source at maximum flux.  
Two different methods are selected, namely the median from FLUOR~PS and the DIRECT~PS, noted as $\bar{V}^{2}_{med}$ and $\bar{V}^{2}_{DPS}$, respectively (see Section~\ref{sec:pipeline}). 
Table~\ref{tab:STSvistest} shows the extracted squared visibilities for gradually lowered STS flux. 
The STS calibration source flux was adjusted to match stellar fluxes of different magnitudes as measured on sky with BC-25. The first column indicates the K-band magnitude (K-mag) that the STS flux has been approximated to. Shown are the two types of extracted squared visibilities $\bar{V}^{2}_{med}$ and $\bar{V}^{2}_{DPS}$, with their uncertainties $\sigma_{\upmu}$ and RSE. For a better comparison, the first STS flux is measured with two camera gain settings, to allow comparison with the on-sky measurement as well as a consistent comparison with the rest of this measurement. The camera frame rate is kept constant at 500 fps. % 2025_07_28_NOSTAR: fld_001 - 004

\begin{table*}
    \centering
        \caption{Laboratory reference measurements using the maximum STS power as calibration source, comparison of the squared visibilities at different days.}
    \label{tab:STSvis}
    \begin{tabular}{ccccccccc}
         Date (UT)  & $\bar{V}^{2}_{med}$ & $\sigma$ & $\sigma_{\upmu}$ & RSE (\%) &  $\bar{V}^{2}_{DPS}$ & $\sigma$ & $\sigma_{\upmu}$ & RSE (\%)\\
         \hline
         2024-05-02 & 0.747 & 0.035 & 0.002 & 0.33 & 0.753 & 0.031 & 0.002 & 0.29  \\ % N = 200 (DPS) / 200 (Fl), fld 002 (bin = 4) (RedFluor: 2024_05_02/)
         \hline
         2024-10-22 & 0.86 & 0.064 & 0.005 & 0.53 & 0.879 & 0.023 & 0.002 & 0.19  \\ % N = 200 (DPS) / 200 (fl), fld 003 (bin = 4) (RedFluor: 2024_10_22/)
         \hline
         2025-07-22 & 1.003 & 0.016 & 0.001 & 0.11 & 1.017 & 0.017 & 0.001 & 0.12  \\ % N = 200 (DPS) / 200 (fl), fld 008 (bin = 1) (RedFluor: 2025_07_23/)
         2025-07-23 & 1.007 & 0.027 & 0.002 & 0.22 & 1.005 & 0.053 & 0.004 & 0.43  \\ % N = 150 (DPS) / 150 (FluorPS), fld 001 (RedFluor: 2025_07_24/)
         2025-07-24 & 1.031 & 0.02 & 0.001 & 0.13 & 1.039 & 0.033 & 0.002 & 0.22  \\% N = 200 (DPS) / 200 (fl), fld 002 (bin = 1) (RedFluor: 2025_07_25/)
         2025-07-26 & 1.044 & 0.03 & 0.002 & 0.20 & 1.039 & 0.039 & 0.003 & 0.27  \\ % N = 200 (DPS) / 200 (fl), fld 001 (RedFluor: 2025_07_26/)
         2025-07-27 & 1.049 & 0.018 & 0.001 & 0.12 & 1.055 & 0.021 & 0.001 & 0.14  \\ % N = 200 (DPS) / 200 (fl), fld 002 (RedFluor: 2025_07_27/)
    \end{tabular}
\end{table*}

\begin{table*}
    \centering
        \caption{Laboratory measurements using the STS calibration source with flux settings that are comparable to stellar fluxes of different magnitude (first column), where on-sky measurements with BC-25 were taken as references.  % 2025_07_28_NOSTAR: fld_001 - 004
    }
    \label{tab:STSvistest}
    \begin{tabular}{cccccccccc}
         STS flux & G$_{APD}$ & $\bar{V}^{2}_{med}$ & $\sigma$ & $\sigma_{\upmu}$ & RSE (\%) &  $\bar{V}^{2}_{DPS}$ & $\sigma$ & $\sigma_{\upmu}$ & RSE (\%)\\
         \hline
         K-mag 2   & 30 & 0.97 & 0.085 & 0.006 & 0.62 & 0.971 & 0.061 & 0.004 & 0.44 \\ % '2025_07_28_NOSTAR_fld_001.fit', N(med) = 200, N(dps) = 200 (bin=1)
         K-mag 2   & 50 & 0.962 & 0.124 & 0.009 & 0.91 & 0.937 & 0.064 & 0.005 & 0.49 \\ % '2025_07_28_NOSTAR_fld_002.fit', N(med) = 199, N(dps) = 200 (bin=1)
         K-mag 3   & 50 & 0.923 & 0.133 & 0.009 & 1.02 & 0.904 & 0.119 & 0.008 & 0.93 \\ % '2025_07_28_NOSTAR_fld_003.fit', N(med) = 199, N(dps) = 200 (bin=1)
         K-mag 4.3 & 50 & 0.75 & 0.32 & 0.03 & 4.06 & 0.85 & 0.32 & 0.02 & 2.71  \\ % '2025_07_28_NOSTAR_fld_004.fit', N(med) = 108, N(dps) = 200 (bin=1)
    \end{tabular}
\end{table*}

\subsection{On-Sky Measurements}
\label{sec:appendixvis}
The extracted squared visibilities of the observation nights during the three engineering runs are presented as an overview in Fig.~\ref{fig:visvsdate}. This section provides additional details and information on the individual results. 

Table~\ref{tab:targets2024A} lists the raw squared visibilities for different targets with magnitudes of around $0 - 4$ during the first engineering run at CHARA with CHARIOT on the 2nd of May 2024 (UT). Shown are the results for $\bar{V}^{2}_{DPS}$ and the associated precisions. As these were the first measurements with CHARIOT, there was yet no complete bracketing sequence (required for visibility calibration and diameter calculation), thus, these results only provide a preliminary understanding of the instrument. Targets for which no usable data were obtained, e.g. due to too low flux or hard- and software issues, are excluded from the table. 

Table~\ref{tab:targets2024B} shows the raw squared visibilities $\bar{V}^{2}_{med}$ and $\bar{V}^{2}_{DPS}$, and their corresponding precisions for three stars with magnitudes ranging between 0.162 and 2.41, which were observed during the second engineering run on the 22nd of October 2024. For each measurement, the name of the target and the seeing at the observed time is included (average of the values measured at the two telescopes).

Table~\ref{tab:targets2025A} lists the raw squared visibilities $\bar{V}^{2}_{med}$ and $\bar{V}^{2}_{DPS}$, and their uncertainties for stellar targets that were observed with CHARIOT during the third engineering run at CHARA in July 2025. 
The target K-band magnitudes range between $\sim 0$ and $\sim 5$. Shown are only a few examples to allow a first comparison with previous observations as well as with the laboratory characterisation.

\begin{table*}
    \centering
        \caption{Overview of the raw squared visibilities $\bar{V}^{2}_{DPS}$, for different targets during the first engineering run at CHARA with CHARIOT on the 2nd of May 2024 (UT).}
        \label{tab:targets2024A}
    \begin{tabular}{c|c|c|c|c|c|c|c|c|}
        \hline 
        \multicolumn{8}{l}{Observation Date: 2024-05-02}\\
        \hline
        Target & K-mag & G$_{APD}$ & Time (UTC) & Seeing (cm) & $\bar{V}^{2}_{DPS}$ & $ \sigma $ & $\sigma_{\upmu}$ & RSE (\%) \\
        \hline
        HD 120933 & 0 & 20 & 6:44 & 8.5  & 0.400 & 0.116 & 0.008 & 2.05 \\ % '2024_05_02_HD_120933_fld_001'
        HD 120933 & 0 & 20 & 6:50 & 8.5 & 0.44 & 0.21 & 0.01 & 3.39 \\ % '2024_05_02_HD_120933_fld_002'
        \hline
        HD 172167 & 0.129 & 20 & 7:40 & 6.6 & 0.50 & 0.15 & 0.01 & 2.1 \\ % '2024_05_02_HD_172167_fld_001'
        HD 172167 & 0.129 & 20 & 7:46 & 6.6 & 0.53 & 0.16 & 0.01 & 2.16\\ % '2024_05_02_HD_172167_fld_002'
        \hline
        HD 164824 & 2.402 & 20 & 7:57 & 8.29 & 0.67 & 0.20 & 0.01 & 2.13\\  % '2024_05_02_HD_164824_fld_001'
        \hline
        HD 181681 & 3.185 & 100 & 10:46 & 12.05 & 1.04 & 0.65 & 0.05 & 4.43\\ % '2024_05_02_HD_181681_fld_001'
        HD 181681 & 3.185 & 100 & 10:52 & 11.59 & 0.99 & 0.62 & 0.04 & 4.41\\ % '2024_05_02_HD_181681_fld_002'
        \hline
        HD 175203 & 3.769 & 100 & 11:05 & 15.79 & 0.99 & 0.74 & 0.05 & 5.26\\ % '2024_05_02_HD_175203_fld_001' 
        \hline
        HD 159410 & 4 & 100 & 12:21 & 15.07 & 1.15 & 0.57 & 0.04 & 3.49 \\ % '2024_05_02_HD_159410_fld_001'
        HD 159410 & 4 & 100 & 12:27 & 15.23 & 1.29 & 0.72 & 0.05 & 3.96 \\ % '2024_05_02_HD_159410_fld_002'
    \end{tabular}

\end{table*}

\begin{table*}
    \centering
        \caption{Overview of the observed targets, seeing conditions, and the measured raw squared visibilities $\bar{V}^{2}_{med}$ and $\bar{V}^{2}_{DPS}$ during the second engineering run at CHARA with CHARIOT on the 22nd of October 2024 (UT).}
        \label{tab:targets2024B}
    \begin{tabular}{c|c|c|c|c|c|c|c|c|c|c|c|c|}
        \hline
        \multicolumn{12}{l}{Observation Date: 2024-10-22} \\
        \hline
         Target & K-mag & G$_{APD}$ & Time (UTC) & Seeing (cm) & $\bar{V}^{2}_{med}$ & $\sigma$ & $\sigma_{\upmu}$ & RSE (\%) &  $\bar{V}^{2}_{DPS}$ & $\sigma$ & $\sigma_{\upmu}$ & RSE (\%)\\
        \hline 
        HD 17506 & 0.162 & 40 & 06:38 & 6.16 &  0.544 & 0.059 & 0.004 & 0.77  & 0.548 & 0.114 & 0.008 & 1.48  \\ % '2024_10_22_HD_17506_fld_001.fit', N(med) = 199, N(dps) = 200 , V2 (aspro) = 0.716
        HD 17506 & 0.162 & 40 & 06:45 & 6.08 & 0.560 & 0.087 & 0.007 & 1.26  & 0.51 & 0.16 & 0.01 & 2.27  \\ % '2024_10_22_HD_17506_fld_002.fit', N(med) = 150, N(dps) = 200, V2 (aspro) = 0.716
        
        HD 17506 & 0.162 & 40 & 07:07 & 6.54 & 0.599 & 0.047 & 0.003 & 0.56  & 0.630 & 0.086 & 0.006 & 0.97  \\ % '2024_10_22_HD_17506_fld_003.fit', N(med) = 196, N(dps) = 200, V2 (aspro) = 0.715
        HD 17506 & 0.162 & 40 & 07:12 & 6.65  & 0.601 & 0.042 & 0.003 & 0.50  & 0.592 & 0.073 & 0.005 & 0.87  \\ % '2024_10_22_HD_17506_fld_004.fit', N(med) = 200, N(dps) = 200, V2 (aspro) = 0.715
        
        HD 17506 & 0.162 & 40 & 07:20 & 7.02 & 0.590 & 0.058 & 0.004 & 0.69  & 0.563 & 0.107 & 0.008 & 1.34  \\ % '2024_10_22_HD_17506_fld_005.fit', N(med) = 200, N(dps) = 200, V2 (aspro) = 0.715
        HD 17506 & 0.162 & 40 & 07:45 & 6.50 & 0.607 & 0.054 & 0.004 & 0.70  & 0.605 & 0.075 & 0.005 & 0.88  \\ % '2024_10_22_HD_17506_fld_006.fit', N(med) = 158, N(dps) = 198, V2 (aspro) = 0.715
        HD 17506 & 0.162 & 40 & 08:16 & 5.62 & 0.594 & 0.054 & 0.004 & 0.67  & 0.562 & 0.089 & 0.007 & 1.17  \\ % '2024_10_22_HD_17506_fld_007.fit', N(med) = 180, N(dps) = 184, V2 (aspro) = 0.717
        \hline
        HD 20902 & 0.54 & 40 & 09:00 & 4.87 & 0.639 & 0.115 & 0.008 & 1.32  & 0.58 & 0.22 & 0.02 & 2.62  \\ % '2024_10_22_HD_20902_fld_001.fit', N(med) = 187, N(dps) = 200, V2 (aspro) = 0.863
        HD 20902 & 0.54 & 40 & 09:24 & 5.90 & 0.64 & 0.13 & 0.01 & 2.02  & 0.59 & 0.18 & 0.01 & 2.12  \\ % '2024_10_22_HD_20902_fld_002.fit', N(med) = 107, N(dps) = 200, V2 (aspro) = 0.864
        \hline
        HD 20123 & 2.41 & 40 & 07:35 & 6.77 & 0.81 & 0.23 & 0.02 & 2.66  & 0.66 & 0.21 & 0.02 & 2.28  \\ % '2024_10_22_HD_20123_fld_001.fit', N(med) = 119, N(dps) = 200, V2 (aspro) = 0.966
        HD 20123 & 2.41 & 40 & 07:58 & 6.54 & 0.78 & 0.23 & 0.02 & 2.08  & 0.61 & 0.55 & 0.04 & 6.38  \\ % '2024_10_22_HD_20123_fld_002.fit', N(med) = 199, N(dps) = 200, V2 (aspro) = 0.966
        
        HD 20123 & 2.41 & 40 & 08:05 & 6.99 & 0.80 & 0.22 & 0.02 & 1.96  & 0.46 & 0.77 & 0.05 & 11.72  \\ % '2024_10_22_HD_20123_fld_003.fit', N(med) = 200, N(dps) = 204, V2 (aspro) = 0.966
        HD 20123 & 2.41 & 40 & 08:32 & 5.33 & 0.88 & 0.21 & 0.02 & 1.86  & 0.61 & 0.58 & 0.04 & 6.73  \\ % '2024_10_22_HD_20123_fld_004.fit', N(med) = 162, N(dps) = 201, V2 (aspro) = 0.966

    \end{tabular}

\end{table*}

\begin{table*}
    \centering
        \caption{Raw squared visibilities $\bar{V}^{2}_{med}$ and $\bar{V}^{2}_{DPS}$, and their uncertainties for several stellar targets during the third engineering run at CHARA with CHARIOT on the 27th of July 2025 (UT).}
    \label{tab:targets2025A}
        \begin{tabular}{c|c|c|c|c|c|c|c|c|c|c|c|c|}
        \hline
        \multicolumn{12}{l}{Observation Date: 2025-07-27} \\
        \hline
         Target & K-mag & G$_{APD}$ & Time (UTC) & Seeing (cm) & $\bar{V}^{2}_{med}$ & $\sigma$ & $\sigma_{\upmu}$ & RSE (\%) &  $\bar{V}^{2}_{DPS}$ & $\sigma$ & $\sigma_{\upmu}$ & RSE (\%)\\
        \hline 
        HD 172167 & 0.129 & 40 & 7:00 & 4.90 & 0.471 & 0.028 & 0.002 & 0.44  & 0.486 & 0.033 & 0.002 & 0.48  \\ % '2025_07_27_HD_172167_fld_001.fit', N(med) = 192, N(dps) = 200, UTC = 07:00, V2 (aspro) = 0.558
        HD 172167 & 0.129 & 40 & 7:27 & 5.61 & 0.474 & 0.038 & 0.003 & 0.57  & 0.490 & 0.041 & 0.003 & 0.58  \\ % '2025_07_27_HD_172167_fld_002.fit', N(med) = 194, N(dps) = 200, UTC = 07:27, V2 (aspro) = 0.566
        HD 172167 & 0.129 & 30 & 8:46 & 2.41 & 0.509 & 0.044 & 0.004 & 0.76  & 0.476 & 0.134 & 0.009 & 1.99  \\ % '2025_07_27_HD_172167_fld_003.fit', N(med) = 132, N(dps) = 200,  UTC = 08:46, V2 (aspro) = 0.596
        HD 172167 & 0.129 & 30 & 9:09 & 1.18 & 0.476 & 0.052 & 0.005 & 1.00  & 0.44 & 0.19 & 0.01 & 3.11  \\ % '2025_07_27_HD_172167_fld_004.fit', N(med) = 119, N(dps) = 204,  UTC = 09:09, V2 (aspro) = 0.604

        \hline
        HD 176527 & 2.038 & 30 & 8:57 & 2.34 & 0.785 & 0.099 & 0.008 & 0.96  & 0.71 & 0.14 & 0.01 & 1.43  \\ % '2025_07_27_HD_176527_fld_002.fit', N(med) = 171, N(dps) = 200, UTC = 08:57, V2 (aspro) = 0.847
        HD 210418 & 3.33 & 50 & 10:09 & 3.25 & 0.70 & 0.21 & 0.02 & 3.04  & 0.54 & 0.26 & 0.02 & 3.39  \\ % '2025_07_27_HD_210418_fld_001.fit', N(med) = 102, N(dps) = 205, UTC = 10:09, V2 (aspro) = 0.979
        HD 210860 & 4.355 & 50 & 11:24 & 2.81 & 0.76 & 0.36 & 0.03 & 4.52  & 0.55 & 0.40 & 0.03 & 5.12  \\ % '2025_07_27_HD_210860_fld_001.fit', N(med) = 111, N(dps) = 200, UTC = 11:24, V2 (aspro) = 0.985
        HD 210824 & 5.212 & 50 & 12:00 & 2.95 & 0.80 & 0.52 & 0.04 & 4.61 & - & - & - & - \\ % '2025_07_27_HD_210824_fld_001.fit', N(med) = 196, UTC = 12:00, V2 (aspro) = 0.99
    \end{tabular}
\end{table*}

% Don't change these lines
\bsp	% typesetting comment
\label{lastpage}
\end{document}